\documentclass[preprint,aps,12pt,notitlepage,nofootinbib,tightenlines]{revtex4}
\usepackage{amsmath}
\usepackage{enumerate} 
\usepackage{bm}
\usepackage{braket}
\usepackage{color}
\usepackage{slashed}
\usepackage{multirow}
\usepackage{array}
\usepackage{float}
\usepackage{feynmp-auto}
\usepackage{booktabs}
\usepackage{threeparttable}
\usepackage{appendix}
\usepackage{ulem}

\usepackage{graphicx}
\definecolor{nicered}{rgb}{0.7,0.1,0.1}
\definecolor{nicegreen}{rgb}{0.1,0.5,0.1}
\usepackage[pagebackref]{hyperref}
\hypersetup{colorlinks=true,citecolor=nicegreen,linkcolor=nicered,urlcolor=magenta}

\newcommand{\be}{\begin{equation}\begin{aligned}}
\newcommand{\ee}{\end{aligned}\end{equation}}

\begin{document}

\title{Single production of an exotic vector-like $Y$ quark \\
at future high energy $pp$ colliders}

\author{Liangliang Shang$^{1,2}$\footnote{Email: shangliangliang@htu.edu.cn; liangliang.shang@physics.uu.se}, Yuxiao Yan$^1$\footnote{Email: yanyuxiao@stu.htu.edu.cn}, Stefano Moretti$^{2,3}$\footnote{Email: s.moretti@soton.ac.uk; stefano.moretti@physics.uu.se}, Bingfang Yang$^1$\footnote{E-mail: yangbingfang@htu.edu.cn}}

\affiliation{\footnotesize
$^1$School of Physics, Henan Normal University, Xinxiang 453007,
PR China\\
$^2$Department of Physics and Astronomy, Uppsala University, Box 516, SE-751 20 Uppsala, Sweden \\
$^3$School of Physics and Astronomy, University of Southampton, Highfield, Southampton SO17 1BJ, UK
}
\vspace*{-1.0cm}
\begin{abstract}
\vspace*{1cm}
\baselineskip=15pt
{\small
Vector-like quarks have been predicted in various new physics scenarios  beyond the Standard Model (SM). In a simplified modelling  of a $(B,Y)$ doublet including a vector-like quark $Y$, with charge $-\frac{4}{3}$e, there are only two free parameters: the $Y$ coupling $\kappa_{Y}$ and mass $m_Y$. In the five flavor scheme, we investigate the single production of the $Y$ state decaying into $Wb$ at the 
Large Hadron Collider (LHC) Run-III and High-Luminosity LHC (HL-LHC) operating at $\sqrt{s}$ = 14 TeV, the possible High-Energy LHC (HE-LHC) with $\sqrt{s}$ = 27 TeV as well as the Future Circular  Collider in hadron-hadron mode (FCC-hh) with $\sqrt{s}$ = 100 TeV. Through detailed signal-to-background analyses and detector simulations, we assess the exclusion capabilities of the $Y$ state at the different colliders.
We find that this can be improved significantly with increasing collision energy, especially at the HE-LHC and FCC-hh, both  demonstrating an obvious advantage with respect to the HL-LHC 
in the case of high $m_Y$. Assuming a 10\% systematic uncertainty on the background event rate, the exclusion capabilities are summarized as follows:
{(1) the LHC Run-III can exclude the correlated regions of $\kappa_{Y} \in [0.06,0.5]$ and $m_{Y} \in [1500\text{ GeV},3800\text{ GeV}]$ with integrated luminosity $L = 300\text{ fb}^{-1}$;
(2) the HL-LHC can exclude the correlated regions of $\kappa_{Y} \in [0.05,0.5]$ and $m_{Y} \in [1500\text{ GeV},3970\text{ GeV}]$ with $L = 3$ ab$^{-1}$; 
(3) the HE-LHC can exclude the correlated regions of $\kappa_Y \in  [0.06,0.5]$ and $m_{Y} \in [1500\text{ GeV} , 6090\text{ GeV}]$ with $L = 3$ ab$^{-1}$; 
(4) the FCC-hh can exclude the correlated regions of $\kappa_{Y} \in [0.08,0.5]$ and $m_{Y} \in [1500\text{ GeV} , 10080\text{ GeV}]$ with $L = 3$ ab$^{-1}$.}}
\end{abstract}

\maketitle
\newpage
%%====================================================================
\section{introduction}
     In 2012, the ATLAS and CMS experiments at the Large Hadron Collider (LHC) made a significant discovery by confirming the existence of the Higgs boson, thereby providing further validation for the Standard Model (SM) \cite{ATLAS:2012yve,CMS:2012qbp}. However, the SM has certain limits in addressing several prominent issues, such as neutrino masses, gauge hierarchy, dark matter and dark energy.
     %在2012年，ATLAS和CMS实验在大型强子对撞机（LHC）中发现了希格斯玻色子，再次证实了标准模型[1,2]。然而，标准模型也存在一些问题，无法解决一些问题，例如中微子质量、质量层次结构、暗物质和暗能量等。
     In various new physics scenarios like little Higgs models \cite{Arkani-Hamed:2002ikv,Han:2003wu,Chang:2003vs,Cao:2007pv}, extra dimensions \cite{4}, composite Higgs models \cite{Agashe:2004rs,Bellazzini:2014yua,Low:2015nqa,Bian:2015ota,He:2001fz,He:1999vp} and other extended models \cite{6,7,8}, the Vector-Like Quarks (VLQs) are predicted to play a role in resolving the gauge hierarchy problem by mitigating the quadratic divergences of the Higgs field. Such VLQs are fermions with spin  $\frac{1}{2} $ and possess the unique characteristic of undergoing both left- and right-handed component transformations under the Electro-Weak (EW) symmetry group of the SM \cite{9}. Unlike chiral quarks, VLQs do not acquire masses through Yukawa couplings to the Higgs field and therefore have the potential to counterbalance loop corrections to the Higgs boson mass stemming from the top quark of the SM. Furthermore, VLQs can generate characteristic signatures at colliders and have been widely studied (see, for example,  \cite{Banerjee:2023upj, Benbrik:2023xlo, Zeng:2023ljl, Canbay:2023vmj, Belyaev:2023yym, Shang:2023ebe, Yang:2023wnv, Bhardwaj:2022nko, Bhardwaj:2022wfz, Bardhan:2022sif, Shang:2022tkr, Freitas:2022cno, Benbrik:2022kpo, Corcella:2021mdl, VLX2021, Belyaev:2021zgq, Deandrea:2021vje, Dasgupta:2021fzw, King:2020mau, Liu:2019jgp, Benbrik:2019zdp, Xie:2019gya, Bizot:2018tds, Cacciapaglia:2018qep, Cacciapaglia:2018lld, Carvalho:2018jkq, CMS:2018kcw, Barducci:2017xtw, CMS:2017voh, Chen:2016yfv, Arhrib:2016rlj, Cacciapaglia:2015ixa, Angelescu:2015kga, Panizzi:2014dwa, Panizzi:2014tya, Cacciapaglia:2012dd, Okada:2012gy, Cacciapaglia:2011fx, delAguila:1989rq}).
     %其中之一就是类似矢量夸克（VLQs），在许多新的物理模型中，如小希格斯模型[3]、额外维度[4]、复合希格斯模型[5]和其他扩展模型[6-8]，它们被预测用于解决通过调整希格斯的平方发散来解决规范层次结构问题。VLQs（vector-like quarks）可以在强子对撞机中产生特征性的标记，这一问题已经得到广泛研究。
    
     %VLQs是自旋1/2的费米子，在SM（标准模型）电弱对称性群下具有左右手组分变换的相同性质[9]。VLQs不通过与希格斯场的Yukawa耦合获得质量，并且可以抵消SM顶夸克对希格斯玻色子质量的环修正。

A VLQ model typically introduces four new states: $T$, $B$, $X$ and $Y$, their electric charges being $+\frac{2}{3}$, $-\frac{1}{3}$, $+\frac{5}{3}$ and $-\frac{4}{3}$, respectively. In such kind of model, VLQs can be categorized into three types: singlets $(T)$, $(B)$, doublets $(X, T)$, $(T, B)$, $(B, Y)$ and triplets $(X, T, B)$, $(T, B, Y)$. {The $Y$ quark does not couple to a SM quark and a SM boson via renormalizable interactions as a singlet.}
However, it is expected to decay with a 100\% Branching Ratio (BR) into a $b$ quark and $W$ boson when $Y$ is lighter than the other VLQs, whether in a doublet or triplet.
 
In this study, we will focus on the observability of single $Y$ production at the Large Hadron Collider (LHC) Run-III, the High-Luminosity LHC (HL-LHC) \cite{Gianotti:2002xx,Apollinari:2017lan}, the High-Energy LHC (HE-LHC) \cite{FCC:2018bvk} and the Future Circular  Collider operating in hadron-hadron mode (FCC-hh) \cite{FCC:2018vvp}, specifically, within the $(B,Y)$ doublet realisation.
    
%类矢量模型是一个假想的模型，新加了四种夸克，命名为：T，B，X， Y。它们的电荷分别是2/3e,-1/3e,5/3e,-4/3e。类矢量夸克有三种形态，单重态[T,B],双重态 [(X,T), (T,B), (B,Y)]，三重态 [(X,T,B), (T,B,Y)]。Y夸克不可能以单重态存在，并且它百分百衰变为Wb。在VLQ模型中，单产生机制是耦合依赖性的，可能在较高的VLQ质量下对总截面有重要贡献。
    
The ATLAS Collaboration conducted a search for single production  of VLQ $Y$ at 13 TeV with an integrated luminosity of 36.1 fb$^{-1}$ \cite{ATLAS:2018dyh}.
They found that the upper limits on the mixing angle are as small as $\left | \sin{\theta_{R}}  \right | $ = 0.17 for a $Y$ quark with a mass of 800 GeV in the $(B, Y)$ doublet model, and $\left | \sin{\theta_{L}}  \right | $ = 0.16 for a $Y$ quark with a mass of 800 GeV in the $(T, B, Y)$ triplet model. 
The CMS Collaboration also conducted a search for single production of $Y$ states in the $W b$ channel at 13 TeV using 2.3 fb$^{-1}$ of data \cite{CMS:2017fpk}. They searched for final states involving one electron or muon, at least one $b$-tagged jet with large transverse momentum, at least one jet in the forward region of the detector plus (sizeable) missing transverse momentum. Their findings indicate that the observed (expected) lower mass limits are 1.40 (1.0) TeV for a VLQ $Y$ with a coupling value of 0.5 and a BR($Y \to W^{-} b$) = 1. 
The ATLAS Collaboration recently presented a search for the pair production of VLQ $T$ in the lepton+jets
final state using 140 fb$^{-1}$ at 13 TeV \cite{ATLAS:2023shy}. They pointed out that
the most stringent limits are set for the scenario BR($T\to W^+ b$)$=1$, for which
$T$ masses below 1700 GeV (1570 GeV) are observed (expected) to be excluded at 95\% Confidence Level (CL).
And the limits can also apply to a VLQ $Y$ with BR($Y\to W^- b$)$=1$.
All such limits stem from VLQ pair production, induced by Quantum Chromo-Dynamics (QCD).

Furthermore, there are comparable exclusion limits on the mixing parameter $\sin{\theta_{R}}$ from EW Precision Observables (EWPOs),
for example within the $(B, Y)$ doublet model, Ref.~\cite{9} found that the upper limits on $\sin{\theta_{R}}$ are approximately 0.21 and 0.15 at $m_Y=1000\text{ GeV}$ and 2000 GeV respectively at 95\% CL from the oblique parameters $S$ and $T$. 
Ref.~\cite{Cao:2022mif} highlighted that, considering the $W$ boson mass measurement by the CDF collaboration \cite{CDF:2022hxs}, the $2\sigma$ bounds on $\sin{\theta_{R}}$ from the oblique parameters $S, T$ and $U$ are approximately $[0.15, 0.23]$ and $[0.09, 0.13]$ at $m_Y=1000\text{ GeV}$ and 3000 GeV in a conservative average scenario, respectively.
They also pointed out that the constraints from the $Zb\bar{b}$ coupling are weaker than those from the EWPOs for about $m_Y>1600\text{ GeV}$.
   
The single production of a VLQ is instead model dependent, as the couplings involved are EW ones, yet they may make a significant contribution to the total VLQ production cross section, compared to the pair production, due to less phase space suppression, in the region of high VLQ masses. {In this work, we will in particular focus on the process $pp\to Y(\to W^- b) j \to l^- \bar\nu_l b j$ in the five flavor scheme (with $l^-$ standing for electron or muon and $j$ standing for first two-generation quark jets), combined with its charged conjugated process $pp \to \bar{Y} j $.} We expect that the forthcoming results will provide complementary information to the one provided by VLQ pair production in the quest to detect a doublet $Y$ quark at the aforementioned future colliders.
	
The paper is structured as follows. In Section \ref{sec:model}, we introduce the simplified VLQ model used in our simulations. In Section \ref{sec:simulation}, we analyze the properties of the signal process and SM backgrounds. Subsequently, we conduct simulations and calculate the $Y$ state exclusion and discovery capabilities at the HL-LHC, HE-LHC and FCC-hh. Finally, in Section \ref{sec:summary}, we provide a summary. 
%\sout{(We also have an Appendix where we map the $Y$ state of our simplified model onto the $(B, Y)$ doublet representation.)}

%本文的组织如下:在第二节，我们介绍了简化的VLQ模型，在第三节，我们分析了信号过程和SM背景的性质。然后，我们进行了仿真，并讨论了第4节的结果。最后，我们做了一个总结。

\section{Doublet VLQ $Y$ in a simplified model}
\label{sec:model}

%VLQ模型是一个假设模型，在其中添加了四个新的夸克，分别为T、B、X和Y。它们的电荷分别为+2/3e、-1/3e、+5/3e和-4/3e。VLQ的单个产生是通过与标准模型夸克耦合实现的。根据标准模型规范群SU(3)C x SU(2)L x U(1)Y，在VLQ中有七种可能的表示，如表I所示。这些表示可以使VLQ与标准模型玻色子和夸克发生耦合。VLQ的动能项和质量项由文献[11]描述。
As mentioned, in a generic VLQ model, one can include four types of states called $T$, $B$, $X$ and $Y$, with electric charges  $+\frac{2}{3}$, $-\frac{1}{3}$, $+\frac{5}{3}$ and $-\frac{4}{3}$, respectively. {Under the SM gauge group, $SU(3)$$_C$ $\times$ $SU(2)$$_L$ $\times$ $U(1)$$_Y$, there are seven possible representations of VLQs as shown in Table \ref{qixingshi}. }

\begin{table}[th]
\begin{tabular}{c|ccccccc}
\hline
VLQ multiplet    & $U$ & $D$  & $Q_1$   & $Q_5$   & $Q_7$   & $T_1$     & $T_2$     \\ \hline
Component fields & $T$ & $B$  & $(T,B)$ & $(B,Y)$ & $(X,T)$ & $(T,B,Y)$ & $(X,T,B)$ \\ \hline
$SU(3)$$_C$      & 3   & 3    & 3       & 3       & 3       & 3         & 3         \\ \hline
$SU(2)$$_L$      & 1   & 1    & 2       & 2       & 2       & 3         & 3         \\ \hline
$U(1)$$_Y$       & 2/3 & -1/3 & 1/6     & -5/6    & 7/6     & -1/3      & 2/3       \\ \hline
\end{tabular}
	\caption{Representations of VLQs and their quantum numbers  under the SM gauge group.}
	\label{qixingshi}
\end{table}

These representations allow for couplings between VLQs and SM gauge  bosons and quarks. The kinetic and mass terms of the VLQs are described as \cite{Cao:2022mif},
\begin{equation}
\label{eq:lagq5}
\mathcal{L}=\sum_{F}^{} \bar{F}(i \slashed{D}-M_{F})F
\end{equation}
where $F =\left \{   U, D, Q_{1}, Q_{5}, Q_{7}, T_{1}, T_{2}\right \}$,
$D_{\mu}=\partial_{\mu}+ig_{1}Y_{F}B_{\mu}+ig_{2}S^{I}W_{\mu}^{I}+ig_{s}T^{A}G^{A}_{\mu}$,
$\lambda^{A}$($A=1,2,\cdots,8$) and $\tau^{I}$($I=1,2,3$), related to the Gell-Mann and Pauli matrices via $T^{A}=\frac{1}{2}\lambda^{A}$ and $S^{I}=\frac{1}{2}\tau^{I}$, respectively. In our simplified model, we use an effective Lagrangian framework for the interactions of a VLQ $Y$ with the SM quarks through $W$ boson exchange, including as $Y$ free parameters $\kappa^{i, L/R}_{Y}$ (couplings) and $m_Y$ (mass) \cite{Buchkremer:2013bha}:
\begin{equation}
   \mathcal{L}= \left\{\kappa_{Y}^{i, L/R} \sqrt{\frac{\zeta_{i} }{\Gamma_{W}^{0} } }\frac{g}{\sqrt{2} }  \left[\bar{Y}_{L/R}W_{\mu}^{-}\gamma^{\mu}d^{i}_{L/R}\right] + \text{H.c.} \right\} + m_Y \bar{Y} Y,
\end{equation}
{where $d^{i}_{L/R}$ represents the three SM generation quarks labeled by $i$, $L$ and $R$ stand for the left-handed and right-handed chiralities, respectively, $\Gamma^0_W = \left(1 - 3 m_W^4/m_Y^4 + 2 m_W^6/m_Y^6 \right)$ for zero SM quark mass $m_q=0$, and $\zeta_i = |V^{4i}_{L/R}|^2/\sum_{i}|V^{4i}_{L/R}|^2$, $V_{L/R}^{4i}$ represents the mixing matrices between the $Y$ quark and the three SM generations.
The current experimental constraints at the LHC tend to favor a $Y$ quark mass at the TeV scale, leading to an approximate value of $\Gamma^{0}_{W} = 1$. We assume that the $Y$ only couples to the SM third  generation quarks, that is, $Y$ decays 100\% into $Wb$ and therefore $\zeta_1=\zeta_2=0, \zeta_3 =1$.
Consequently, the above Lagrangian can be simplified as,}
\begin{equation}
\label{eq:lageff}
   \mathcal{L}= \left\{ \frac{g \kappa_{Y}^{3,L/R}}{\sqrt{2}}  \left[ \bar{Y}_{L/R}W_{\mu}^{-}\gamma^{\mu}b_{L/R} \right] + \text{H.c.} \right\} + m_Y \bar{Y} Y, 
\end{equation}
where $g$ is the EW coupling constant. By comparing the Lagrangian for the $(B, Y)$ doublet and $(T, B, Y)$ triplet, we observe that the relationship between the coupling $\kappa_{Y}^{3,L/R}$ and  mixing angle $\theta^{L/R}$ is $\sin\theta^{L/R} = \kappa_Y^{3,L/R}$ for the doublet and $\sin\theta^{L/R} = \sqrt{2} \kappa_Y^{3,L/R}$ for the triplet.
%, \sout{with details to be found in  Appendix \ref{sec:app}.} 
Taking into account the relationships $\tan\theta^L = \frac{m_b}{m_B}\tan\theta^R$ and $\tan\theta^R = \frac{m_b}{m_B}\tan\theta^L$ as well as the condition $m_B\gg m_b$, we can assume $\kappa_Y^{3,L} = 0$ for the doublet and $\kappa_Y^{3,R}=0$ for the triplet\footnote{In the subsequent content, we will use $\kappa_Y$ to denote $\kappa_Y^{3,R}$ for the sake of simplicity.}.
As mentioned in the introduction, Ref.~\cite{9, Cao:2022mif} found that the upper limits on $\sin\theta_R$ are on the order of $\mathcal{O}(10^{-1})$ at 95\% CL from the EWPOs. Therefore, a value smaller than 0.5 for $\kappa_Y$ is considered in this study. The $Y$ decay width can be expressed as \cite{Cetinkaya:2020yjf},
 \begin{equation}
     \Gamma (Y \to Wq) = \frac{\alpha _{e}\kappa^{2}_{Y}}{16\sin^{2}\theta_{W}} \frac{(m^{2}_{W}-m^{2}_{Y})^{2}(2m^{2}_{W}+m^{2}_{Y})}{m^{2}_{W}m^{3}_{Y}},  
 \end{equation}
where $\alpha_{\rm EM} = \frac{{g^\prime}^2}{4\pi}$, $g^\prime$ is the Electro-Magnetic (EM) coupling constant and $\theta_W$ is the EW  mixing angle.
In this paper, we solely focus on the Narrow Width Approximation (NWA), which we use for the purpose of simplifying scattering amplitude calculations. However, it is worth noting that several studies \cite{Carvalho:2018jkq, Berdine:2007uv, Moretti:2016gkr, Deandrea:2021vje} have highlighted the limitations of the NWA in scenarios involving new physics with VLQs. Specifically, it becomes imperative to consider a finite width when this becomes larger than $\alpha_{\rm EM}\approx 1\%$, given the substantial interference effects emerging between VLQ production and decay channels, coupled with their interactions with the corresponding irreducible backgrounds. %Furthermore, when dealing with larger widths, the exclusion range tends to expand. 
To address the limitations of our approach then, we will also present the ratio $\Gamma_Y/m_Y$ in our subsequent results and emphasise since now that, crucially, for the region where $\Gamma_Y/m_Y > 1\%$, our sensitivities may be under- or over-estimated, as such interferences could be positive or negative, respectively.
Besides, 
before starting with our numerical analysis, we remind the reader that one can apply the results of our forthcoming simulations to a  specific VLQ representation, such as, e.g.,  $(B, Y)$ or $(T, B, Y)$, by utilizing the aforementioned relationships.

{There are stringent limits from the oblique parameters $S$, $T$ and $U$ from the EWPOs \cite{Hollik:1988ii, Peskin:1990zt, Grinstein:1991cd, Peskin:1991sw, Lavoura:1992np, Burgess:1993mg, Maksymyk:1993zm, Cynolter:2008ea, 9, Chen:2017hak, Cao:2022mif, He:2022zjz, Arsenault:2022xty}.
	These oblique parameters relate to weak isospin currents and 
	EM currents, involving their vacuum polarization amplitudes. Consequently, the oblique parameters can be reformulated using the vacuum polarizations of the SM gauge bosons.
	The contributions in the doublet $(B, Y)$ model to these oblique parameters can be approximated as follows~\cite{Cao:2022mif}:}
\begin{equation}
S \simeq \frac{1}{2\pi}\left\{-\frac{2}{3}\kappa_Y^2 \ln\frac{\mathcal{M}^2}{m_b^2}+\frac{11}{3}\kappa_Y^2\right\},\,
U \simeq - \frac{\kappa_Y^2}{2\pi}, \,
T \simeq \frac{3m_t^2}{8\pi \sin^2\theta_W m_W^2}\kappa_Y^4 \frac{2\mathcal{M}^2}{3m_t^2}.
\end{equation}
Here, $\mathcal{M}^2 = (m_Y^2-m_b^2\kappa_Y^2)/(1-\kappa_Y^2)$ and $m_W=m_Z\cos\theta_W$. For the numerical calculation, the $\chi^2$ function for the oblique parameter fit should be less than 8.02 for three degrees of freedom to compute the $2\sigma$ limits, yielding:  ${S}=-0.02\pm0.1$, ${T}=0.03\pm0.12$, ${U}=0.01\pm0.11$. Note that there exists a strong correlation of 92\% between the $S$ and $T$ parameters while the $U$ parameter exhibits an anti-correlation of $-80\% (-93\%)$ with $S$ ($T$) ~\cite{ParticleDataGroup:2022pth}. Specific numerical values of the input parameters are detailed in Eq.~(\ref{eq:input}).

\begin{figure}[!htb]
	\begin{center}
		\includegraphics[scale=0.45]{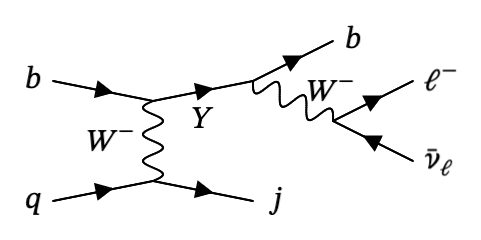}
		\caption{{Representative Feynman diagram of single $Y$ production $pp\to Y j$ followed by its subsequent decay $Y\to W^-(\to l^- \bar \nu_l) b$ in the five flavor scheme. Here, $q$ in the initial state represents one of the first two generation quarks, {$j$ in the final state represents one of the first two generation jets}, $b$ in the initial (final) state represents a $b$-quark (jet) while $l^-$ represents either an electron or muon.} 
  %\sout{Notice that, since we use the five flavor scheme, the $g\to b\bar b$  splitting in the diagram is actually accounted for through the PDF evolution.}
  }
		\label{fm}
	\end{center}
\end{figure}

%VLQs are exclusively coupled to SM quarks and Higgs fields through the Yukawa interaction, and they are commonly produced in association with other SM particles. 
%VLQ（矢量样夸克）只与标准模型夸克和希格斯场之间通过Yukawa相互作用耦合，并通常与其他标准模型粒子独立产生。产生截面不仅取决于VLQ质量，还取决于VLQ与标准模型夸克之间的耦合强度。
In Figure~\ref{fm}, we show a representative Feynman diagram of the signal production $pp\to Y j$ and decay chain $Y\to W^-(\to l^- \bar\nu_l) b$. We expect the $W$ boson and the high-momentum $b$-jet to exhibit a back-to-back alignment in the transverse plane, originating from the decay of the massive $Y$ quark. The topology also encompasses an outgoing light quark, often resulting in a forward jet within the detector. %\sout{Furthermore, the second $b$-jet arising from the splitting of a gluon into a pair of $b$-quarks can be observed in either the forward or central region.} 
{According to these features of signal events, the primary SM backgrounds include $pp \to t(\to l^+ b \nu_l)j$, $pp \to W^\pm(\to l \nu_l)bj$, $pp \to W^\pm(\to l \nu_l)W^\mp(\to j j) b$ plus their charge conjugated processes.
Amongst these, the first two are irreducible backgrounds while the last one is a reducible background. We have also assessed additional backgrounds, such as $pp \to t\bar{t}$ {and $pp \to Zbj$}, and found that their contributions can be ignored based on the selection criteria that will be discussed later.}
 \begin{figure}[htbp]
	\centering	
	\includegraphics[width=11cm]{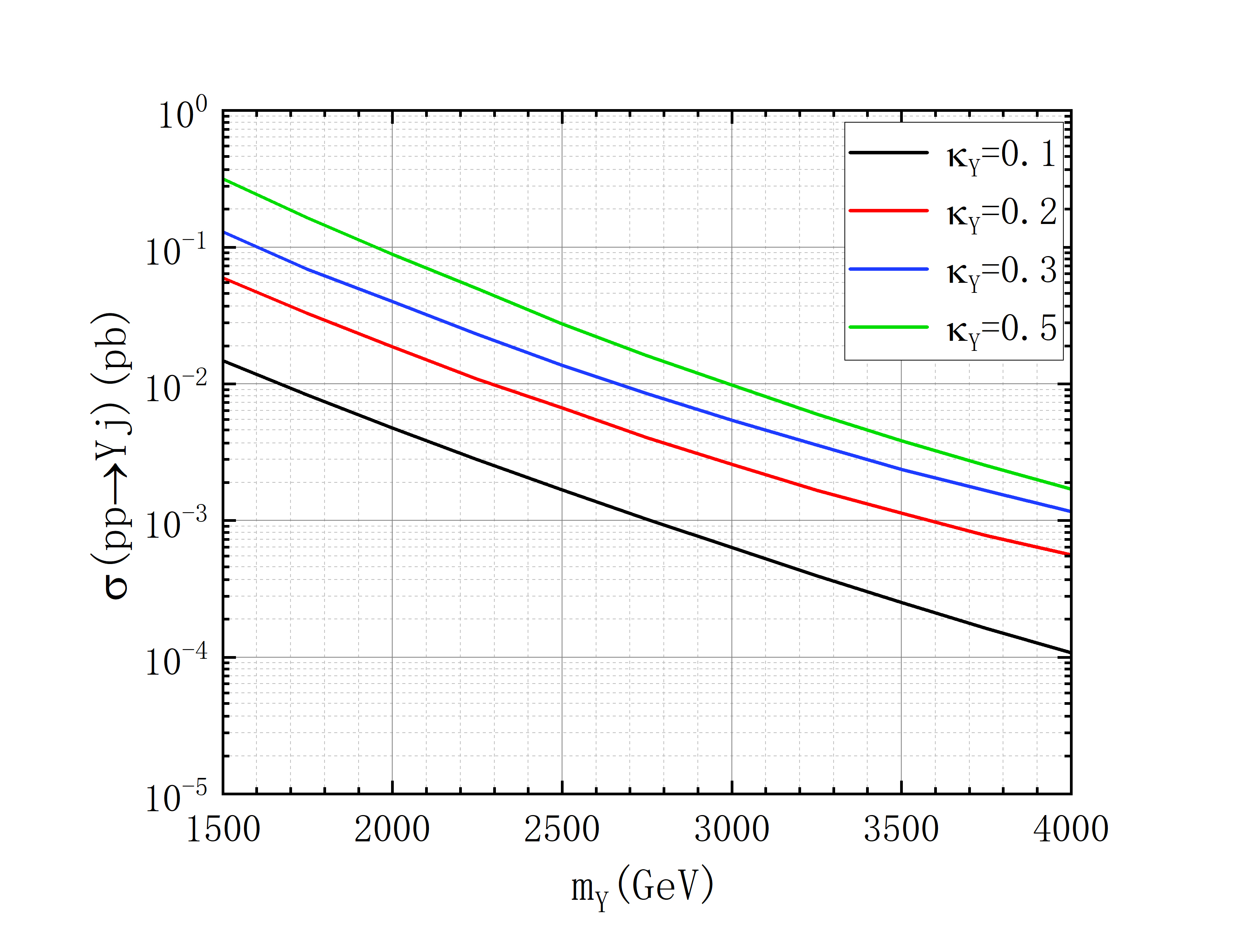}\vspace{-0.5cm}
	\caption{The tree-level cross sections for single $Y$ production $pp\to Y(\to W^- b) j$ as a function of the mass $m_Y$ for various values of the coupling $\kappa_{Y}$. The charge conjugated process has also been included.} 
	\label{ky} %此处的label相当于一个图片的专属标志，目的是方便上下文的引用
\end{figure}

The signal production cross section is determined not only by the mass $m_Y$ but also by the coupling strength $\kappa_Y$. 
%图2中可以找到代表性的费曼图，描述了产生和衰变链。需要注意的是，分析中还包括共轭荷过程。
The cross section is directly proportional to  $\kappa_{Y}^{2}$ for a fixed $m_Y$ as long as the NWA is met \cite{Moretti:2016gkr}.
In Figure~\ref{ky}, we show the tree-level cross sections for  single $Y$ production as a function of the mass $m_Y$. {We can see that, as $m_Y$ increases, the cross section gradually decreases. Actually, this is mostly  attributed to the fact that the gluon Parton Distribution Function (PDF) grow smaller at the larger $x$ values needed to produce the ever more massive final state.}
%随着V LQ−Y夸克质量的增加，由于更小的相空间，单个VLQ-Y产生的截面逐渐减小，如图1所示。

%According to the article on the mass of the W boson \cite{Cao:2022mif}, it is observed that in the case of vector-like $Y$ quarks, the triplet $(T, B, Y)$ has a cross section that is twice as large as that of the doublet $(B, Y)$. In this study, our focus is on investigating the properties of the doublet $(B, Y)$.
%根据《W玻色子质量》文章的研究结果，我们发现对于类矢Y夸克来说，三重态（T、B、Y）的产生截面是双重态（B、Y）的两倍。很明显，对于给定的类矢Y夸克质量，单个VLQ-Y的产生截面与其耦合强度Y的平方成正比。在本文中，我们专注于研究双重态（B、Y）的性质。
\begin{figure}[htbp]
	\centering	
	\includegraphics[width=9cm]{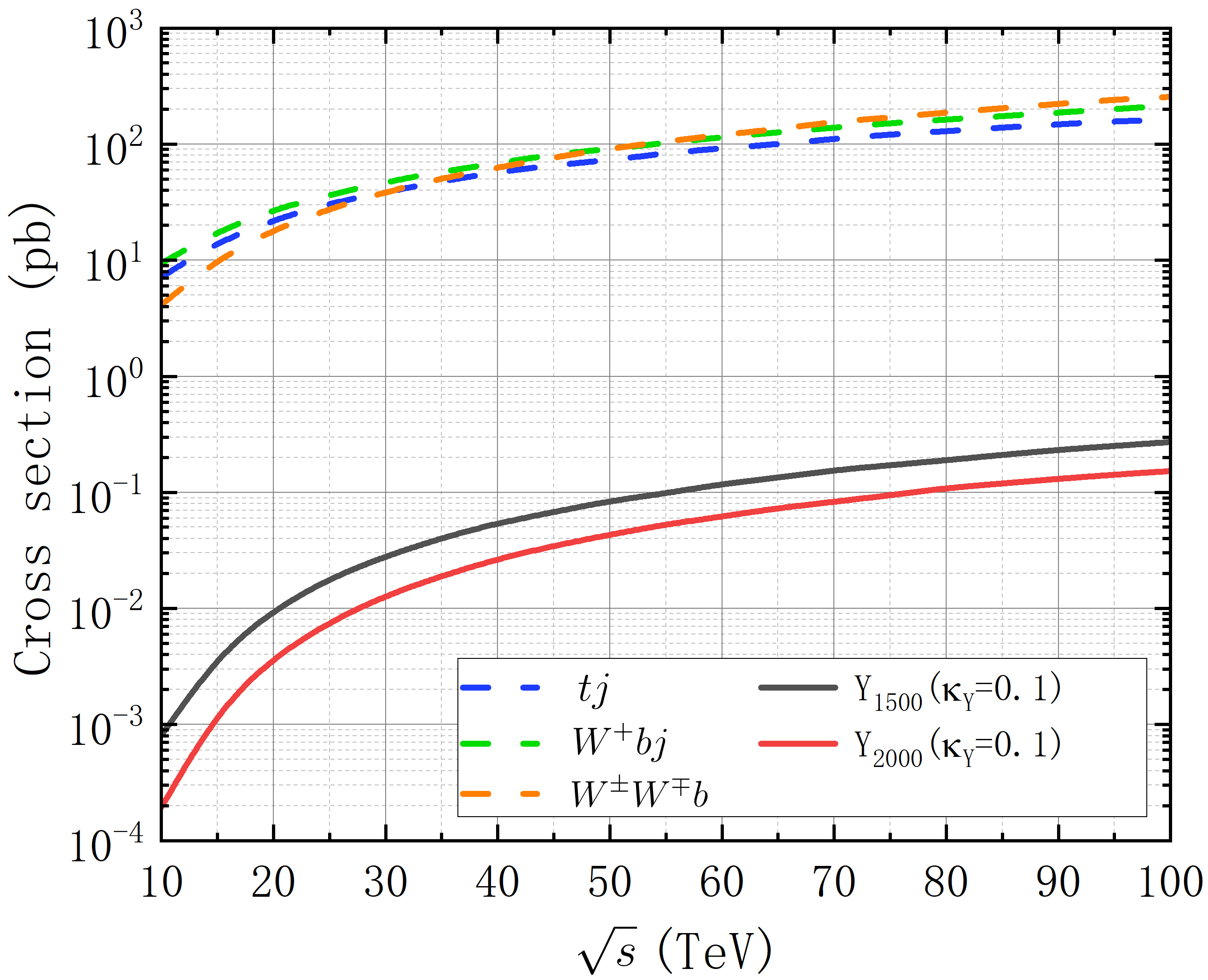}\vspace{-0.5cm}
	\caption{The tree-level cross sections as a function of the center-of-mass energy $\sqrt{s}$ for the signal benchmarks and backgrounds. Solid lines represent the signal processes  $pp\to Y(\to l^- \bar\nu_l b) j$ and dashed lines represent the background processes $pp \to t(\to l^+ b \nu_l)j$, $pp \to W^\pm(\to l \nu_l)bj$, $pp \to W^\pm(\to l \nu_l)W^\mp(\to j j) b$. The cross sections also include the charge conjugated processes. } 
	\label{jiemian}
\end{figure}

{In Figure~\ref{jiemian}, we show the tree-level cross sections for the signal benchmarks $m_Y=1500\text{ GeV}$ (labeled as $Y_{1500}$) and $m_Y=2000\text{ GeV}$ (labeled as $Y_{2000}$) with $\kappa_{Y} =0.1$ as well as the tree-level cross sections for the background processes. The signal cross sections for different values of $\kappa_Y$ can be obtained by rescaling.} It is evident that the rates for the backgrounds are significantly larger than those for the signals. Consequently, we should design efficient selection criteria in terms of kinematic cuts to reduce the number of background events while preserving the signal events. Furthermore, the cross sections for both signal and backgrounds increase with increasing collider energy.
%, reflecting variations across different colliders such as the HL-LHC, HE-LHC and FCC-hh. With same integrated luminosity, it is more advantageous for us to design selection criteria when the signal cross sections are sufficiently large before implementing cuts.

The Next-to-Leading Order (NLO) or even higher order QCD corrections for the SM background cross sections at the LHC have been extensively explored in Refs. \cite{deBeurs:2018pvs, Kidonakis:2018ncr, Boos:2012vm, Campbell:2006cu}. The $K$ factors associated with the background cross sections adopted in our calculations are summarized in Table~\ref{Kyinzi}\footnote{Note that, despite they change somewhat with energy, we neglect here the changes of $K$ factors values at different colliders, like in Ref.~\cite{Yang:2022wfa}.}. 
%式中\delta是对背景的测量时不可避免出现的不确定度这里，主要背景的QCD修正是通过包含一个K因子来考虑的，如T able.II所示。

\begin{table}[H]
\centering
\begin{tabular}{ccccc}
\hline
%Processes & $Zbj$    & $W^{+}bj$   & $W^{+}W^{-}b$  & $t\bar{b}j$ \\ \hlin e
%$K$ factor  & 1.3 \cite{Campbell:2005zv}      & 1.9 \cite{Campbell:2006cu}         &2.1 \cite{Campbell:2006cu}           &  1.4 \cite{Kidonakis:2018ncr,Boos:2012vm} \\ \hline
Processes & $tj$    & $W^{\pm}bj$   & $W^{\pm}W^{\mp}b$  \\ \hline
$K$ factor  & 1.1 \cite{deBeurs:2018pvs,Kidonakis:2018ncr,Boos:2012vm}      & 1.9 \cite{Campbell:2006cu}         &2.1 \cite{Campbell:2006cu} \\ \hline
\end{tabular}
\caption{$K$ factors representing the QCD corrections for the background processes.}
\label{Kyinzi}
\end{table}

\section{Signal to background analysis}   
\label{sec:simulation}
 The signal model file is sourced from \text{FeynRules} \cite{feynruls} and parton-level events are generated using \text{MadGraph5\_aMC$@$NLO} \cite{Alwall:2014hca} with the PDF of NNPDF23LO1 \cite{NNPDF}. 
Dynamic factorization and renormalization scales are set as default in \text{MadEvent} \cite{website_factor}.
Subsequently, fast detector simulations are conducted using \text{Delphes 3.4.2} \cite{deFavereau:2013fsa} with the built-in detector configurations of the LHC Run-III, HL-LHC, HE-LHC \cite{website_hllhc} and FCC-hh \cite{website_fcchh}. Jets are clustered by \text{FastJet} \cite{Cacciari:2011ma}  employing the anti-$kt$ algorithm \cite{Cacciari:2005hq} with a distance parameter of $\Delta R = 0.4$. Furthermore, \text{MadAnalysis 5} \cite{Conte:2012fm} is used to analyze both signal and background events. Finally, the \text{EasyScan\_HEP} package \cite{Shang:2023gfy} is utilized to connect these programs and scan the VLQ parameter space.
 %信号模型文件来自FeynRules \cite{feynruls}，使用NN23LO1 \cite{NNPDF} 部分子分布函数（PDF）使用MadGraph5_aMC@NLO生成部分子级事件。然后，这些事件传递给PYTHIA 8 [66]进行喷注和强子化。使用DELPHES 3.4.2 [67]进行快速探测器模拟，模拟了HL-LHC、HE-LHC和FCC-hh的探测器配置。使用FastJet [68]和anti-kt算法 [69]以及R = 0.4的距离参数进行喷注聚类。最后，使用MadAnalysis 5 [70]来分析信号和背景事件。为了连接这些程序并扫描参数空间，使用EasyScan_HEP [71]软件包。

The numerical values of the input SM parameters are taken as follows~\cite{ParticleDataGroup:2022pth}:
\begin{align}
 m_b = 4.18{\rm ~GeV},\quad m_t = 172.69{\rm ~GeV},\quad m_{Z} =91.1876 {\rm ~GeV}, \notag \\
 \sin^{2}\theta_W = 0.22339,\quad \alpha(m_Z) =1/127.951, \quad \alpha_s(m_Z)=0.1179.
\label{eq:input}
\end{align}
{Considering the general detection capabilities of detectors, the following basic cuts at the parton level are chosen\footnote{{Note that $\Delta R>0.4$ is required as part of the parton-level settings in run\_card.dat of \text{MadGraph5\_aMC$@$NLO}. This requirement is implemented to enhance the percentage of events generated at the parton level that satisfy the criteria for isolated particles. For events involving isolated particles in \text{MadAnalysis 5}, we similarly impose $\Delta R > 0.4$ at the detector simulation level in delphes\_card.dat, following the default requirement in \text{Delphes}.}}:}
\begin{equation}
	\begin{aligned}
	    \Delta R(x,y)>0.4\ \ \ \ \ \ &(x,y=l,j,b),\\ \nonumber
		 p^l_T>25 \ {\rm GeV}, \ \ \ \ \ \ \ &\ \ |\eta_l|<2.5, \\
		p^j_T>20 \ {\rm GeV}, \ \ \ \ \ \ \ &\ \ |\eta_j|<5.0, \\
		p^b_T>25 \ {\rm GeV}, \ \ \ \ \ \ \ &\ \ |\eta_b|<2.5, \\
	\end{aligned}
\end{equation}
where $\Delta R = \sqrt{\Delta \Phi  ^{2}+ \Delta  \eta ^{2} } $ denotes the separation in the rapidity($\eta$)–azimuth($\phi$) plane. 

%%%%%%%%%%%%%%%%%%%%%%%%%%%%%%%%%%%%%%%%%%%%%%%%%
%数据统计与分析
%%%%%%%%%%%%%%%%%%%%%%%%%%%%%%%%%%%%%%%%%%%%%%%%%
To handle the relatively small event number of signal ($s$) and background ($b$) events, we will use the median signiﬁcance $\mathcal{Z}$ to estimate the expected discovery and exclusion reaches  \cite{Cowan:2010js, Kumar:2015tna},
%为了处理相对较小的事件数，我们将使用中位数显著性SS来估计预期的发现显著性和排除显著性[20]背景事例的模拟过程中不可避免地将产生源自于各种不确定性的系统不确定度。系统不确定度的主要来源有四个方面：
  %（1）亮度与截面的不确定性
  %（2）探测器相关的实验不确定性
  % (3)数据驱动的对背景估计的不确定性
  % (4)模拟过程中的建模不确定性
  \begin{align}   \label{eq:excl}
      \mathcal{Z}_{excl}
      =\sqrt{2\left[s-b\ln\left(\frac{b+s+x}{2b} \right)-\frac{1}{\delta ^{2}}\ln\left(\frac{b-s+x}{2b} \right) \right]-(b+s-x)\left(1+\frac{1}{\delta^{2}b } \right)},
  \end{align}   

  \begin{align}    \label{eq:disc}
    \mathcal{Z}_{disc}=\sqrt{2\left[(s+b)\ln\left(\frac{(s+b)(1+\delta^{2}b) }{b+(s+b)\delta ^{2}b} \right)-\frac{1}{\delta ^{2}}\ln\left( 1+\frac{\delta ^{2}s}{1+\delta ^{2}b} \right) \right]},
  \end{align}   

   \begin{equation}
       x=\sqrt{(s+b)^{2}-\frac{4\delta ^{2}sb^{2}}{1+\delta ^{2}b} }, 
   \end{equation}
%During the simulation of the background case, various uncertainties inevitably lead to system uncertainties. There are four primary sources of systematic uncertainty, that is, the uncertainty in luminosity and cross section, the uncertainty stemming from detector-related experiments, the uncertainty in data-driven background estimation and the modeling uncertainty during simulation.
where $\delta$ is the uncertainty that inevitably appears in the measurement of the background. 
The exclusion capability corresponds to $\mathcal{Z}_{\text{excl}}=2$ while the discovery potential corresponds to $\mathcal{Z}_{\text{disc}}=5$.
In the completely ideal case, that is $\delta$=0, Eq.~(\ref{eq:excl}) and (\ref{eq:disc}) can be simplified as follows:
 \begin{equation}
     \mathcal{Z}_{excl}=\sqrt{2\left[s-b\ln\left(1+\frac{s}{b} \right)\right]},
 \end{equation}
 and
 \begin{equation}
     \mathcal{Z}_{disc}=\sqrt{2\left[(s+b)\ln\left(1+\frac{s}{b} \right)-s\right]}.
 \end{equation}

\subsection{LHC Run-III and HL-LHC}

%%%%%%%%%%%%%%%%%%%%%%%%%%%%%%%%%%%%%%%%%%%%%%
%cut分析
%%%%%%%%%%%%%%%%%%%%%%%%%%%%%%%%%%%%%%%%%%%%%%
%信号与背景的区别
Firstly, we establish a {selection} that emulates the LHC Run-III and HL-LHC detector response based on the count of final state particles detected in each event. Given the limited efficiency of the detector in identifying jets, we adopt a lenient approach towards the number of jets. {Consequently, the final count selection is defined as follows: $N_{l} = 1$, $N_{b} \ge 1$, $N_{j} \ge 1$.

Considering that the mass of $Y$ is notably greater than that of its decay products, the latter exhibit distinct spatial characteristics in  pseudorapidity $\eta$ and spatial separation $\Delta R$ compared to backgrounds. These differences inform our selection criteria.

\begin{figure*}[h!]
	\centering
	\begin{minipage}{0.47\linewidth}
		\centering
		\includegraphics[width=1.1\linewidth]{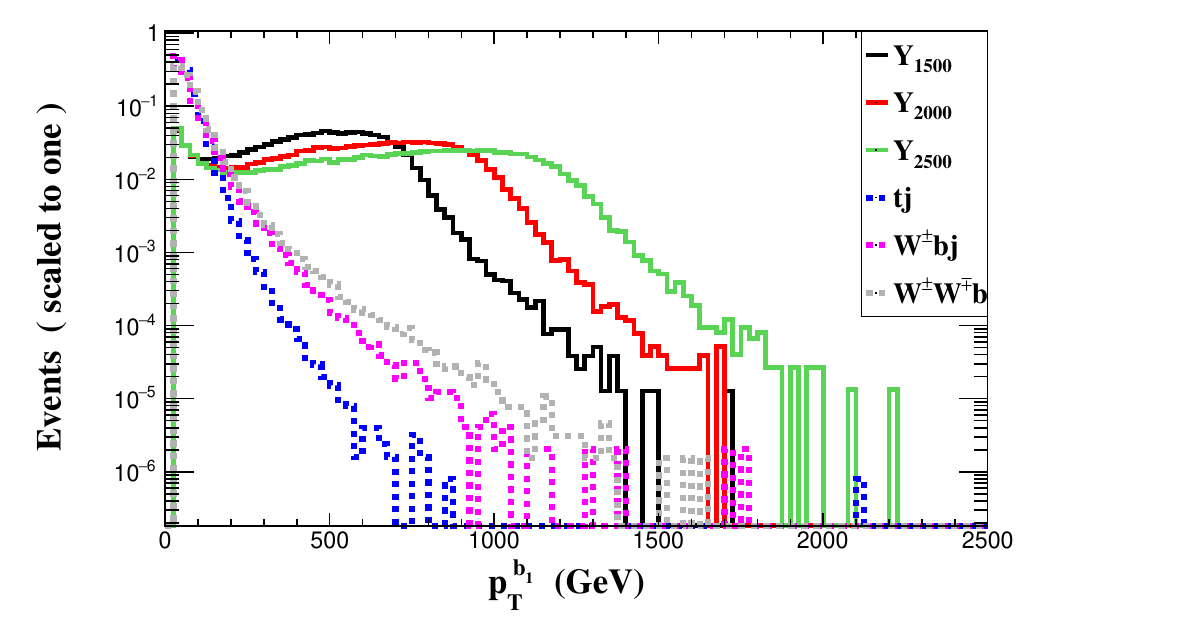}
		%\label{chutian1}%文中引用该图片代号
	\end{minipage}
	%\qquad
	\begin{minipage}{0.47\linewidth}
		\centering
		\includegraphics[width=1.1\linewidth]{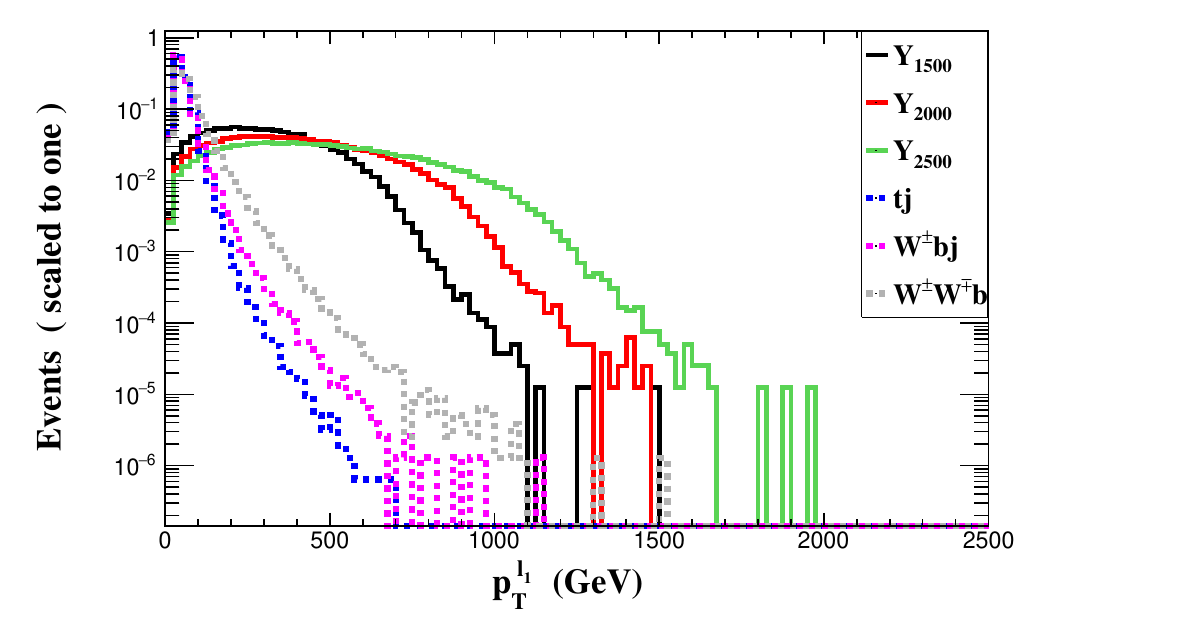}
	\end{minipage}
	\begin{minipage}{0.47\linewidth}
		\includegraphics[width=1.1\linewidth]{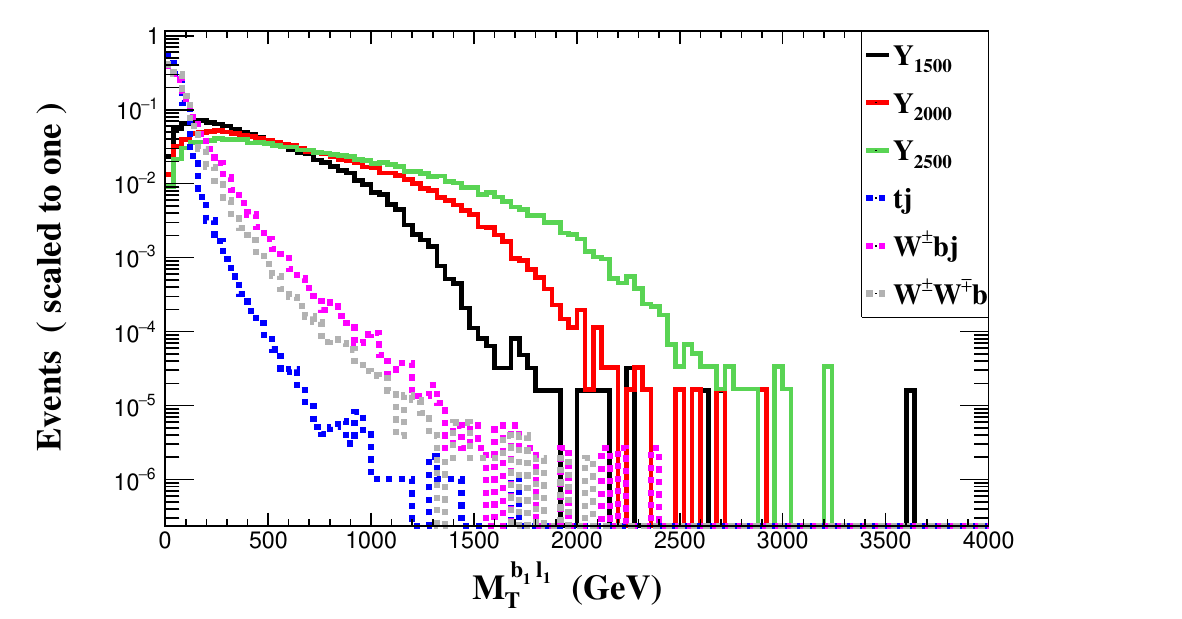}
	\end{minipage}   
	\begin{minipage}{0.47\linewidth}
		\includegraphics[width=1.1\linewidth]{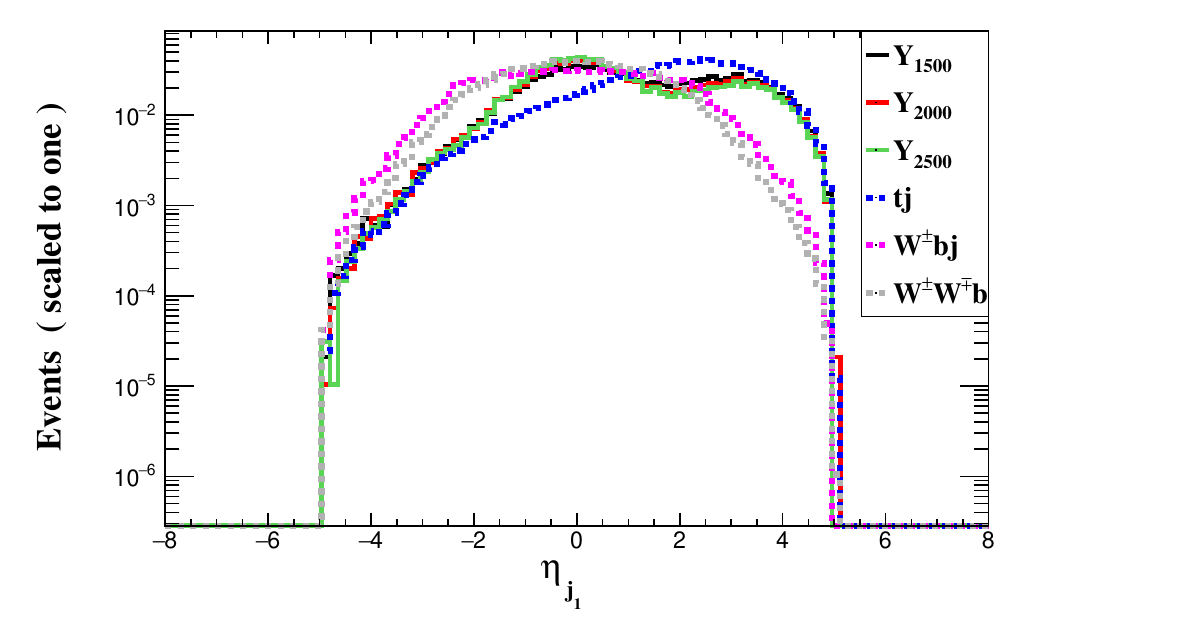}
	\end{minipage}
	%\qquad
	\begin{minipage}{0.47\linewidth}
		\centering
		\includegraphics[width=1.1\linewidth]{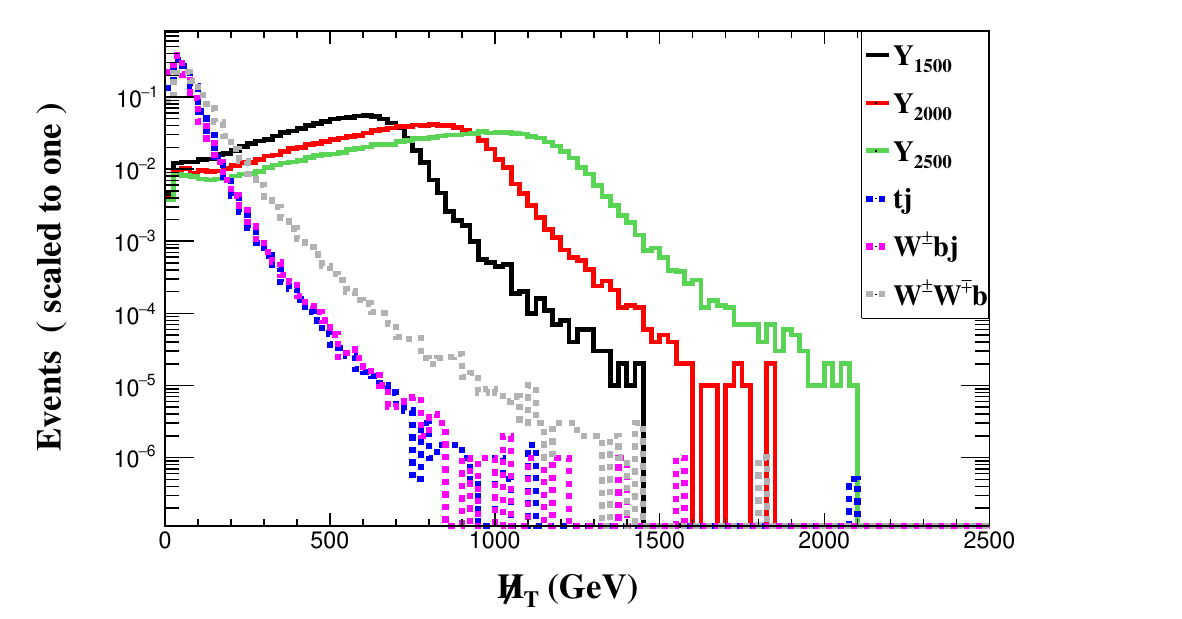}
	\end{minipage}
        \begin{minipage}{0.47\linewidth}
		\centering
		\includegraphics[width=1.1\linewidth]{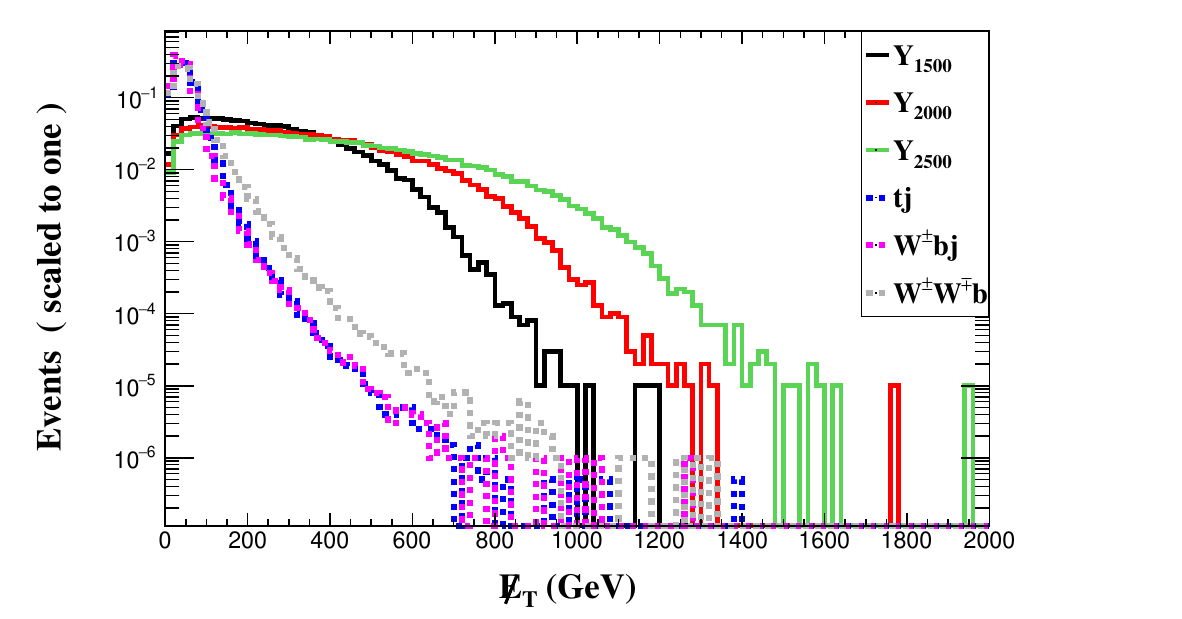}
	\end{minipage}
	\caption{ Normalized distributions for the signals of $m_Y=$ 1500 GeV, 2000 GeV and 2500 GeV and SM backgrounds at the HL-LHC.
		The conjugated processes have been included.
	} 
	\label{zhifangtu}
\end{figure*}

Furthermore, 
since the mass range of $Y$ is much heavier than the particles originating from background processes, we anticipate that the transverse momentum (referred to as $\vec{p}_T$ and its magnitude denoted as $p_T$) of decay products of the $Y$ state will be substantially larger than those of the same particles from the background processes. Besides, we will also consider variables such as $\slashed{E}_T$, $\slashed{H}_T$, $M_T(b_1, l_1)$, $M(b_1, l_1)$ and $M(b_1, l_1, j_1)$ to distinguish the signal from the background.
Here, {$\slashed{E}_T$ is missing energy representing the magnitude of the sum of the transverse momenta of all visible final state particles, $\slashed{H}_T$ is analogous to $\slashed{E}_T$ but only considers all visible hadronic momenta}, $M$ is reconstructed mass by energy-momentum of specific final states, while the transverse mass $M_T$ is defined as follows:
\begin{align*}
    M_T^2 & \equiv [E_T(1)+E_T(2)]^2-[\vec{p}_T(1)+\vec{p}_T(2)]^2 \\
    & = m_1^2 + m_2^2 + 2 [E_T(1)E_T(2) - \vec{p}_T(1)\cdot \vec{p}_T(2)],
\end{align*}
where $E_T(i)=\sqrt{p^2_T(i)+m^2_i}$ and $m^2_i = p_i^2$ with $p_i$ representing a 4-vector.

In Figure~\ref{zhifangtu}, we present the normalized distributions of $p_{T}^{b_{1}}$, $p_{T}^{l_{1}}$, $M_{T}^{b_1 l_1}$, $\eta_{j_1}$, $\slashed{H}_{T}$ and $\slashed{E}_{T}$ for  $m_Y=1500\text{ GeV}$, $m_Y=2000\text{ GeV}$ and $m_Y=2500\text{ GeV}$ with $\kappa_Y=0.1$ as well as for the background processes. Based on these distributions, we have devised the following selection criteria to distinguish the signal from the various backgrounds \footnote{The subscript on the particle symbol is arranged according to the magnitude of the particle transverse momentum: e.g., in the case of $b$-jets, $p_T^{b_1}$ is greater than $p_T^{b_2}$.}:
%在本节中，我们将研究未来14 TeV HL-LHC上的矢量型Y夸克的信号。我们在图3中展示了两个信号基准点（Y1500，Y1800）以及√s = 14 TeV的背景下Pj1T、Pb1T、Pl1T、MblT和H/T的归一化分布。根据这些分布的特点，我们采用复杂的切割条件以增强信号的显著性。

\begin{itemize}
    \item {Cut-1}: $N_{l} = 1$, $N_{b} \ge 1$, $N_{j} \ge 1$;
    \item Cut-2: $p_{T}^{b_{1}} > 350\text{ GeV}$ and $p_{T}^{l_{1}} > 200\text{ GeV}$; 
   % \item Cut-3: $M_{b_1 l_1} > 500\text{ GeV}$; 
   % \item Cut-4: $M_{b_1 l_1 j_1} > 700\text{ GeV}$;
    \item {Cut-3}: $M_{T}^{b_1 l_1} > 300\text{ GeV}$;
    \item Cut-4: $\eta_{j_1} > 1.8$; 
    \item {Cut-5}: $\slashed{H}_{T} > 300\text{ GeV}$ and $\slashed{E}_{T} > 100\text{ GeV}$.
\end{itemize}

By applying these cuts, we can see that the signal efficiencies for $m_Y=1500\text{ GeV}$, $m_Y=2000\text{ GeV}$ and $m_Y=2500\text{ GeV}$ are approximately 8\%, 12\% and 13\%, respectively. The higher efficiency for the latter can be attributed to the larger transverse boost of the final state originating from an heavier $Y$. Meanwhile, the background processes are significantly suppressed. For reference, we provide the cut flows in Table \ref{cutflow14}.
%通过应用这些削减，mY = 1500 GeV和mY = 1800 GeV的信号效率分别为1.35％和2.41％。后者的效率更高，这可以归因于来自较重Y的最终状态的横向动量增强更显著。同时，背景过程被显着抑制。供参考，每个削减级别的截面在表III.1中提供。

\begin{table}[ht]
	\centering
	\begin{threeparttable}
		\begin{tabular}{c|ccc|c|c|c}
			\hline
			Cuts & $Y_{1500}$ (fb) & $Y_{2000}$ (fb) & $Y_{2500}$ (fb)   &$tj$\,\, (fb)  &  $W^{\pm}bj$\,\,(fb)    &$W^\pm W^\mp b$\,\,(fb)     \\ \hline
			Basic Cuts   & 3.06     & 0.98  & 0.35   & 14117     & 30001     & 18967   \\
			Cut 1        & 1.70     & 0.54  & 0.19   & 5773.50   & 8379.80   & 8635.20  \\
			Cut 2        & 0.83     & 0.32  & 0.12   & 0.07      & 5.85      & 12.61   \\
			%Cut 3        & 0.83     & 0.32  & 0.12   & 0.06      & 5.80      & 11.34   \\
			%Cut 4        & 0.83     & 0.32  & 0.12   & 0.05      & 5.61      & 11.09   \\
			Cut 3        & 0.51     & 0.25  & 0.10   & 0.03      & 3.18      & 7.08    \\
			Cut 4        & 0.28     & 0.13  & 0.05   & 0.01      & 0.21      & 0.48    \\
			Cut 5        & 0.24     & 0.12  & 0.05   & 0.01      & 0.09      & 0.27    \\ \hline
		\end{tabular}
	\end{threeparttable}
	\caption{\label{cutflow14} Cut flows of the signal with $\kappa_Y=0.1$ and backgrounds at the 14 TeV HL-LHC, where the conjugate processes $pp \to \bar{t}j$, $W^\mp \bar{b} j$, $W^\mp W^\pm \bar{b}$ have been included.}
\end{table}
%

%%%%%%%%%%%%%%%%%%%%%%%%%%%%%%%%%%%%%%%%%%%%%%%%%%%%%%%%%
%14TeV数据分析
%%%%%%%%%%%%%%%%%%%%%%%%%%%%%%%%%%%%%%%%%%%%%%%%%%%%%%%%%
We present the exclusion capability ($\mathcal{Z}_{\text{excl}}=2$) and discovery potential ($\mathcal{Z}_{\text{disc}}=5$) for $Y$ with two different integrated luminosities, 1000 fb$^{-1}$ and 3000 fb$^{-1}$, at the HL-LHC, as shown in the top line of Figure~\ref{14-liangdutu}. 
{In detail, we commence the calculation from $m_{Y}=1500\text{ GeV}$ up to 4000 GeV and select benchmark points at intervals of 250 GeV with $\kappa_Y=0.1$ to obtain their cross sections ($\sigma_s$) and efficiencies ($\epsilon_s$) for the process $pp\to Y j \to W^-(\to l^- \bar\nu_l) b j$. For each signal benchmark, we generate $10^5$ events, and for each background, we generate $10^6$ events. We have verified that different values of $\kappa_Y$ yield the same efficiency when $m_Y$ remains constant. Furthermore, as illustrated in Figure~\ref{fm}, the cross section $\sigma_s$ is approximately proportional to the square of $\kappa_Y$. Consequently, we can set $s \approx L \times \sigma_s / (0.1)^2 \times \kappa_Y^2 \times \epsilon_s$ in Eq.~(\ref{eq:excl}) and (\ref{eq:disc}). Thus, for a specific integrated luminosity ($L$), we can determine the corresponding $\kappa_Y$ value when $Y$ is discovered or excluded.}
We consider both the ideal scenario without systematic uncertainties and the case with a 10\% systematic uncertainty.
In the presence of 10\% systematic uncertainty, the $Y$ can be excluded in the correlated parameter space of $\kappa_{Y} \in [0.06, 0.5]$ and $m_{Y} \in [1500\text{ GeV}, 3800\text{ GeV}]$ with an integrated luminosity of $L=300\text{ fb}^{-1}$, which corresponds to the maximum achievable integrated luminosity during LHC Run-III.
If the integrated luminosity is raised to 3000 fb$^{-1}$, aligning with the maximum achievable at the HL-LHC, the excluded parameter space extend to $\kappa_{Y} \in [0.05, 0.5]$ and $m_{Y} \in [1500\text{ GeV}, 3970\text{ GeV}]$.
%Furthermore, the discovery regions are $\kappa_{Y} \in [0.072 , 0.5]$ ([0.047, 0.5]) and $m_{Y} \in [1000\text{ GeV}, 2621\text{ GeV}]$ ($[1000\text{ GeV}, 3047\text{ GeV}]$) with $L=300\text{ fb}^{-1}$ (3000 fb$^{-1}$).

\subsection{27 TeV HE-LHC}

\begin{figure*}[h!]
	\centering
	\begin{minipage}{0.47\linewidth}
		\centering
		\includegraphics[width=1.1\linewidth]{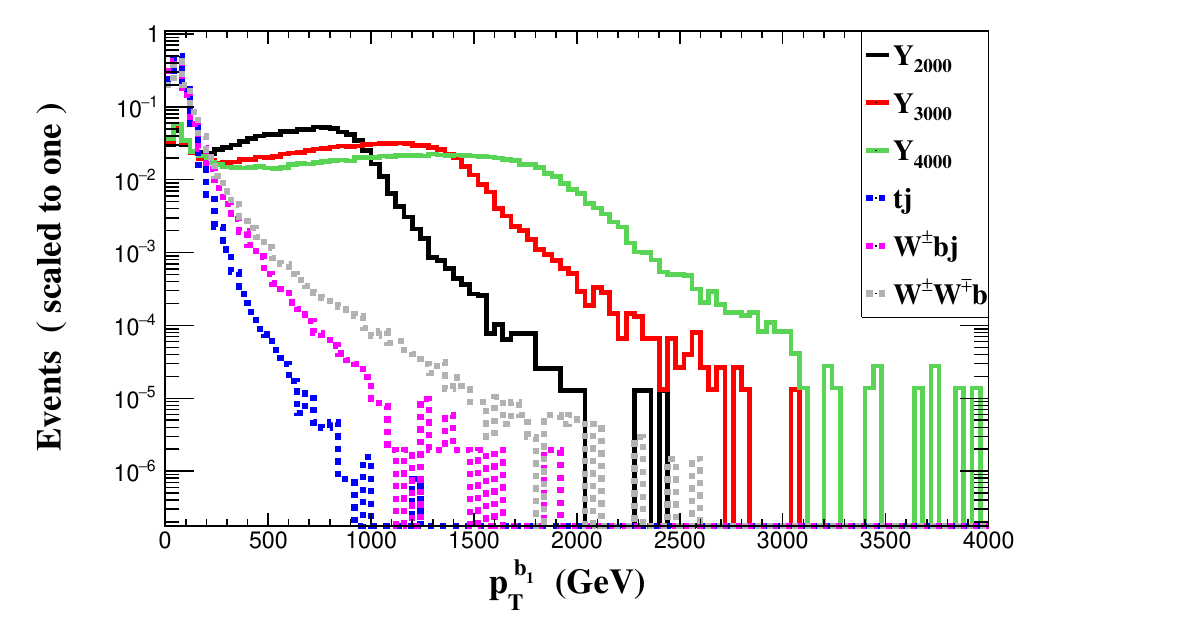}
		%\label{chutian1}%文中引用该图片代号
	\end{minipage}
	%\qquad
	\begin{minipage}{0.47\linewidth}
		\centering
		\includegraphics[width=1.1\linewidth]{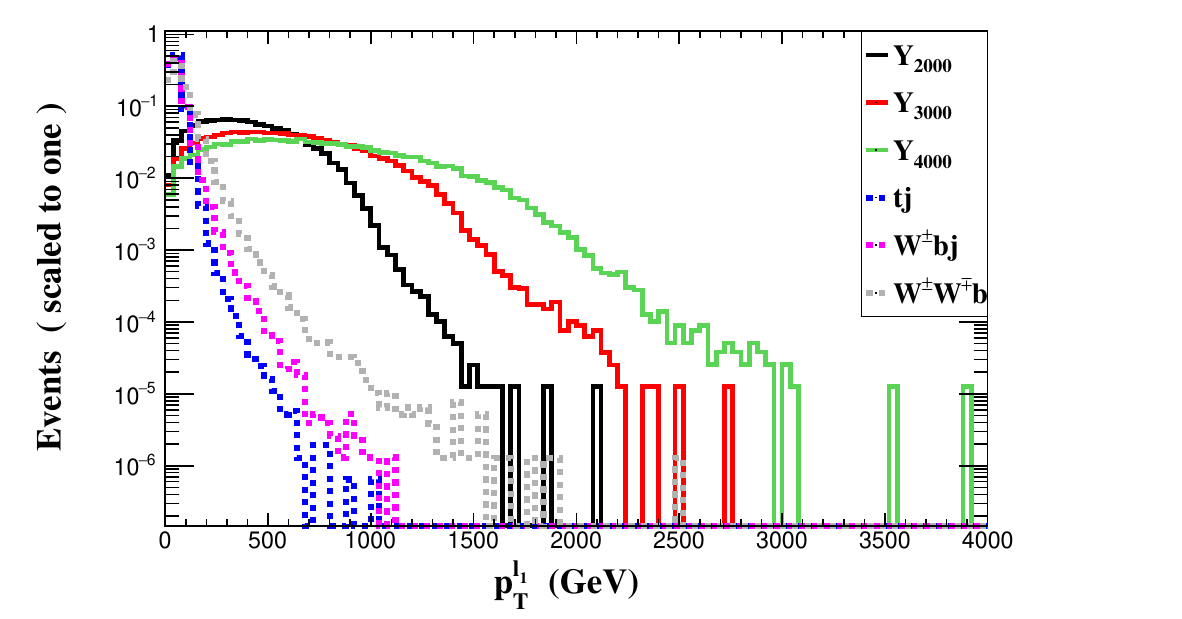}
	\end{minipage}
	\begin{minipage}{0.47\linewidth}
		\includegraphics[width=1.1\linewidth]{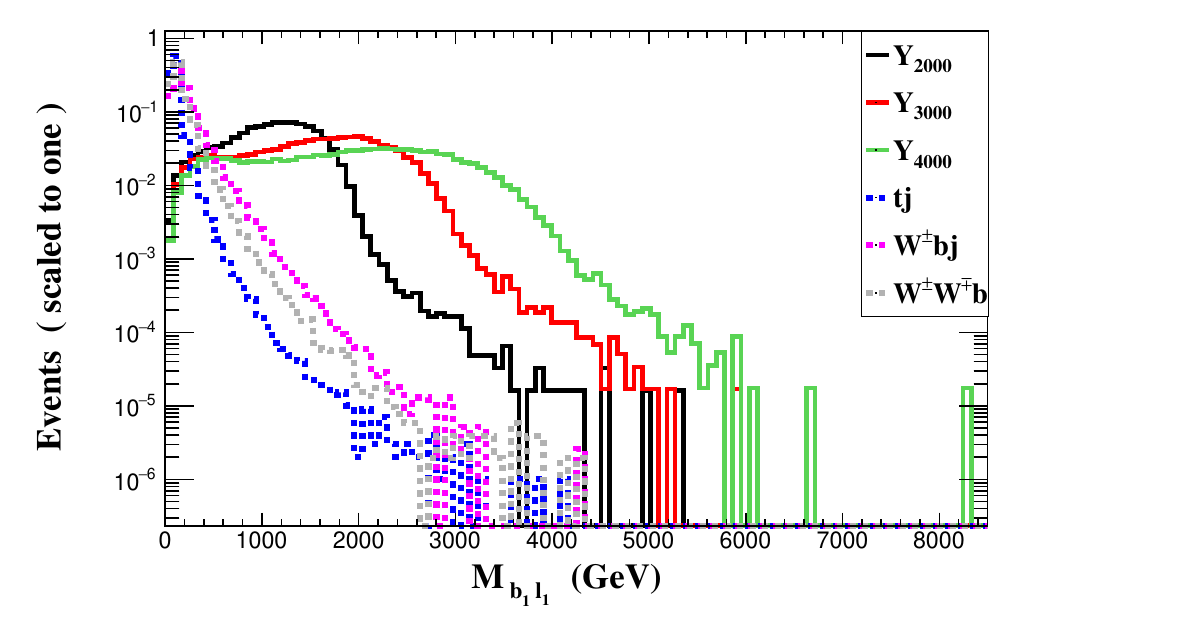}
    \end{minipage}
    \begin{minipage}{0.47\linewidth}
		\includegraphics[width=1.1\linewidth]{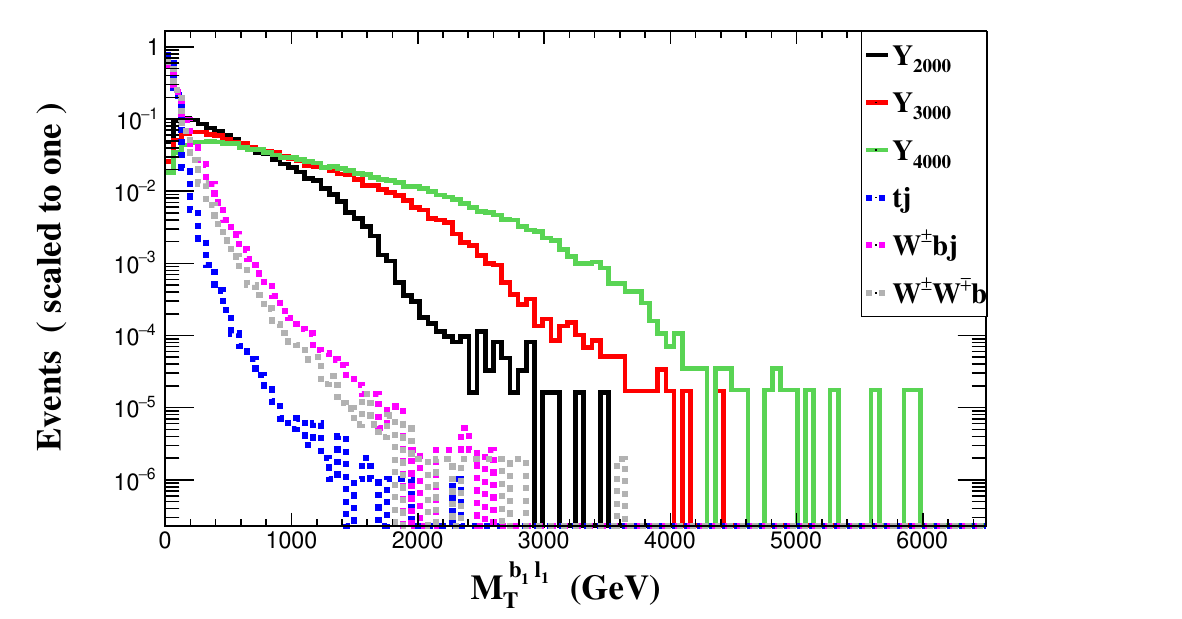}
		%\label{chutian1}%文中引用该图片代号
    \end{minipage}    
 \begin{minipage}{0.47\linewidth}
		\includegraphics[width=1.1\linewidth]{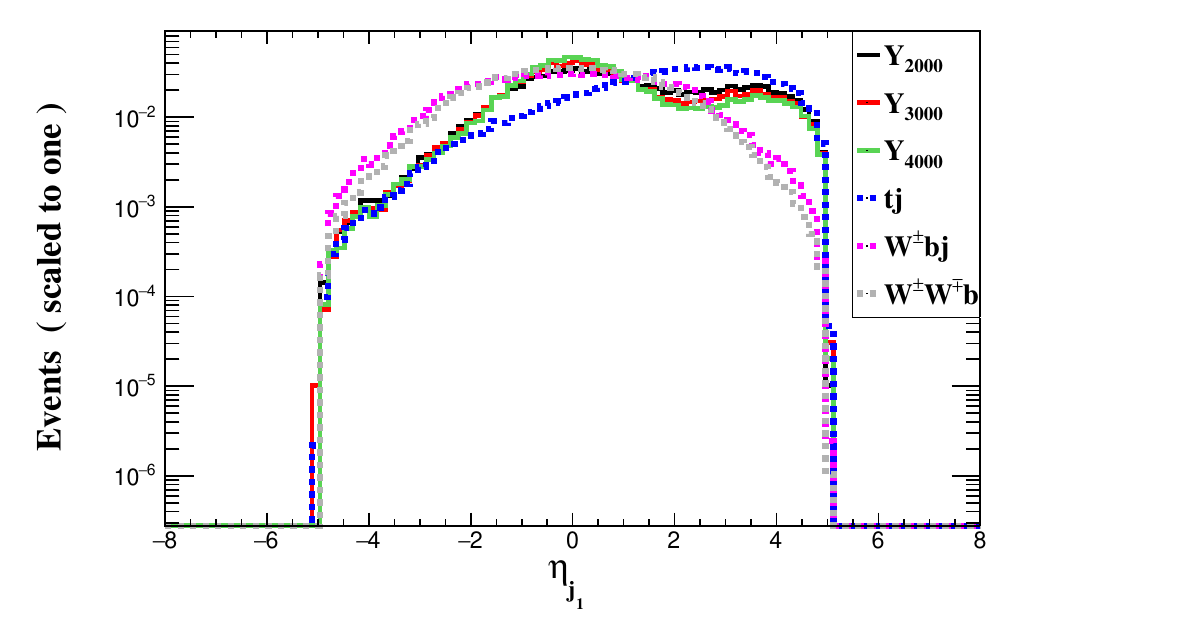}
	\end{minipage}
	%\qquad
	\begin{minipage}{0.47\linewidth}
		\centering
		\includegraphics[width=1.1\linewidth]{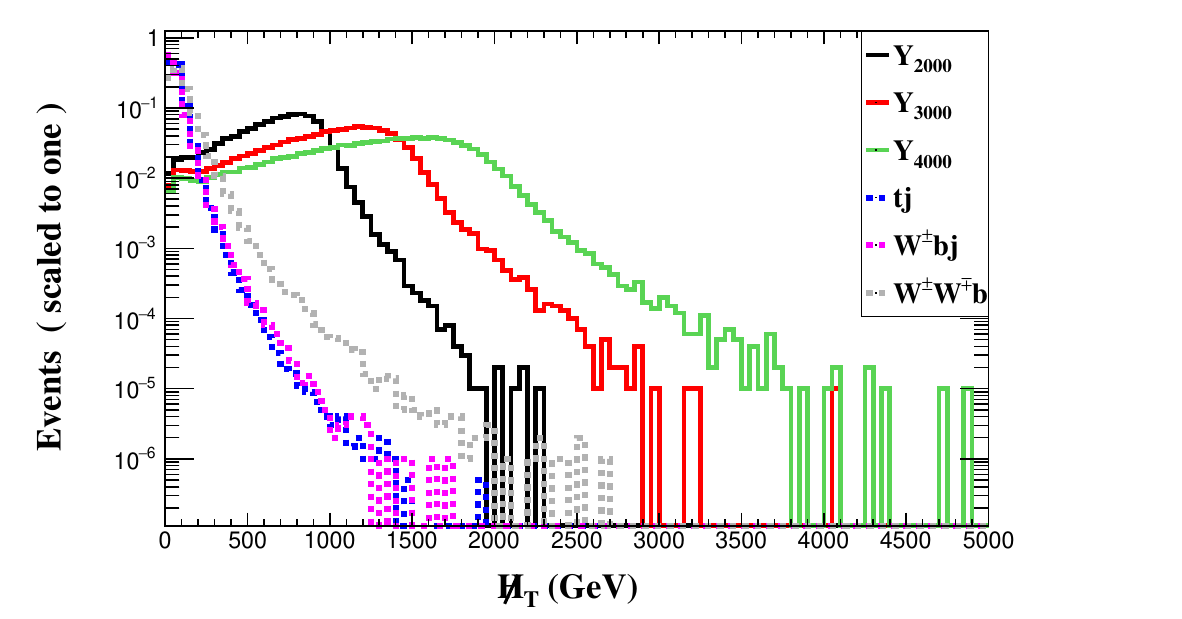}
	\end{minipage}
        \begin{minipage}{0.47\linewidth}
		\centering
		\includegraphics[width=1.1\linewidth]{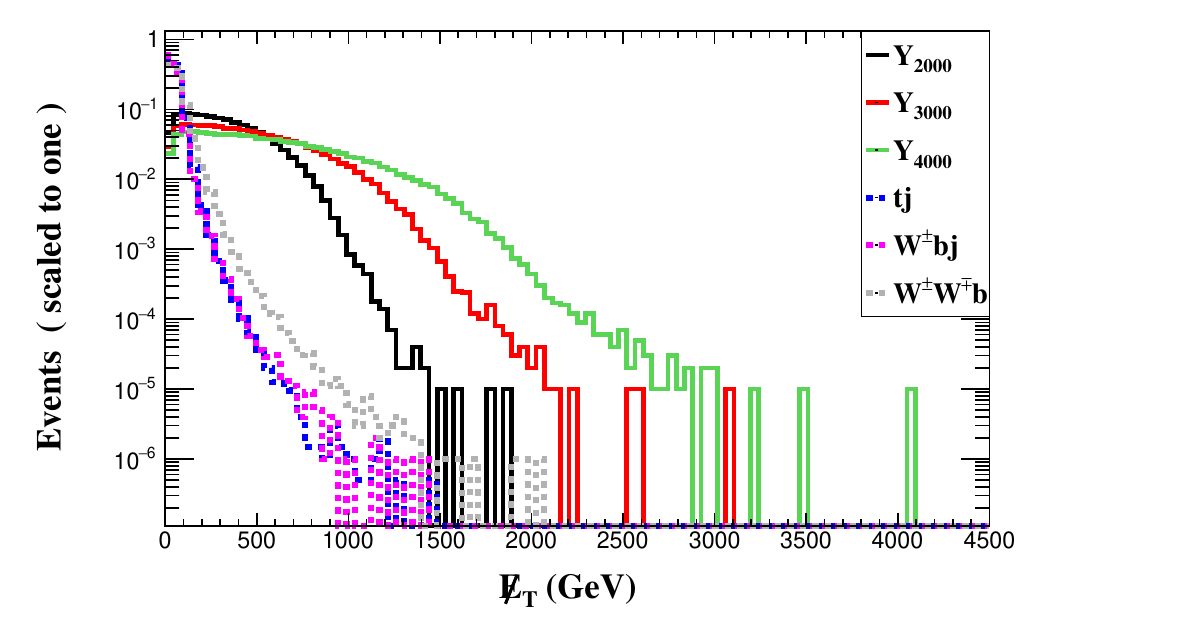}
	\end{minipage}
	\caption{ Normalized distributions for the signals with $m_Y=$ 2000 GeV, 3000 GeV and 4000 GeV and backgrounds at the HE-LHC.} %The conjugated processes have been included.} 
		\label{27fenbutu}
\end{figure*}

%信号与背景的区别

This section delves into the prospective signal of $Y$ at the future 27 TeV HE-LHC. We commence the calculation from $m_{Y}=1500\text{ GeV}$ up to 7000 GeV and select benchmark points at intervals of 250 GeV with $\kappa_Y=0.1$.
In Figure~\ref{27fenbutu}, we exhibit the normalized distributions for signal of $m_Y = 2000\text{ GeV}, 3000\text{ GeV}$ and 4000 GeV as well as background processes, forming the basis for our distinctive selection criteria:
%在本节中，我们将研究未来27TeV HE-LHC上矢量型Y夸克的信号。图5展示了两个信号基准点（Y1500、Y1800）以及√s = 27TeV背景下PT [j1]、PT b1、PT l1、MT [b1l1]和H/T的归一化分布。根据这些分布的特征，我们采用复杂的切割条件以增强信号的显著性。

    \begin{itemize}
    \item {Cut-1}: $N_{l} = 1$, $N_{b} \ge 1$, $N_{j} \ge 1$;
    \item Cut-2: $p_{T}^{b_1} > 300 \text{ GeV}$ and $p_{T}^{l_1} > 250 \text{ GeV}$; 
    \item Cut-3: $M_{b_1 l_1} > 1000 \text{ GeV}$; 
   % \item Cut-4: $M_{b_1 l_1 j_1} > 900 \text{ GeV}$;
    \item {Cut-4}: $M_T^{b_1 l_1} > 2.0 \text{ GeV}$; 
    \item Cut-5: $\eta_{j_1} > 2.0$; 
    \item {Cut-6}: $\slashed{H}_{T} > 400 \text{ GeV}$ and $\slashed{E}_{T} > 100 \text{ GeV}$.
\end{itemize}
The kinematic variables compared to the 14 TeV case, an additional variable $M_{b_1 l_1}$ is introduced here. The cut threshold values for transverse momentum-based variables, such as $\slashed{H}_{T} > 400 \text{ GeV}$, are higher than those in the 14 TeV case. This adjustment accounts for the increased center-of-mass energy. Detailed cut flows are outlined in Table \ref{cutflow27} and the exclusion capability and discovery potential are shown in the second row of Figure~\ref{14-liangdutu}. The signal efficiencies for $m_Y$ = 2000 GeV, $m_Y$ = 3000 GeV and $m_Y$ = 4000 GeV are approximately 9\%, 13\% and 13\%, respectively. 
The $Y$ quark can be excluded within the correlated parameter space of $\kappa_{Y} \in [0.06, 0.5]$ and $m_{Y} \in [1500\text{ GeV}, 6090\text{ GeV}]$ with 10\% systematic uncertainty for $L= 10 $ ab$^{-1}$. We note that the loss of sensitivity at low masses as being due to the cuts being optimised for heavy masses.
%For $L=3000\text{ fb}^{-1}$, the discovery regions are $\kappa_{Y} \in [0.053 , 0.5]$ and $m_{Y} \in [1000\text{ GeV}, 3885\text{ GeV}]$. 
%If the integrated luminosity is raised to the highest designed value 10 ab$^{-1}$, the discovery parameter regions can be extended to $\kappa_{Y} \in [0.051, 0.5]$ and $m_{Y} \in [1000\text{ GeV}, 3943\text{ GeV}]$.

\begin{table}[ht]
	\centering
	\begin{threeparttable}
		\begin{tabular}{c|ccc|c|c|c}
			\hline
			Cuts       & $Y_{2000}$ (fb) & $Y_{3000}$ (fb)& $Y_{4000}$ (fb)  &$tj$\,\, (fb)  &$W^\pm bj$\,\,(fb)    &$W^\pm W^\mp b$\,\,(fb)  \\ \hline
			Basic Cuts   & 9.59   & 2.41 & 0.73 & 37406     & 77274     & 69303   \\
			Cut 1        & 5.34   & 1.29 & 0.38 & 15814.40  & 23225     & 32315     \\
			Cut 2        & 3.03   & 0.86 & 0.26 & 0.56      & 26     & 94.74       \\
			Cut 3        & 2.69   & 0.83 & 0.25 & 0.15      & 11.60     & 25.09        \\
			Cut 4        & 2.07   & 0.74 & 0.24 & 0.07      & 8.19      & 17.88        \\
			Cut 5        & 1.05   & 0.34 & 0.10 & 0.04      & 0.39      & 1.87        \\
			Cut 6        & 0.91   & 0.31 & 0.09 & 0.04      & 0.08      & 1.11        \\ \hline
		\end{tabular}
	\end{threeparttable}
	\caption{\label{cutflow27} Cut flows of the signal with $\kappa_Y=0.1$ and backgrounds at the 27 TeV HE-LHC.}%, where the conjugate processes have been included.}
\end{table}

\subsection{100 TeV FCC-hh}

%我们给出了两个典型信号m的横截面Y = 1500GeV和mY = 1800GeV, T able中实施删减后的相关背景。I One可以看到，所有的SM背景都被非常有效地抑制了，而信号在切割流的末尾仍然有比较好的效率。较大的背景来自pp→W +射流过程，总截面为0.9 nb。
\begin{figure*}[h!]
	\centering
	\begin{minipage}{0.47\linewidth}
		\centering
		\includegraphics[width=1.1\linewidth]{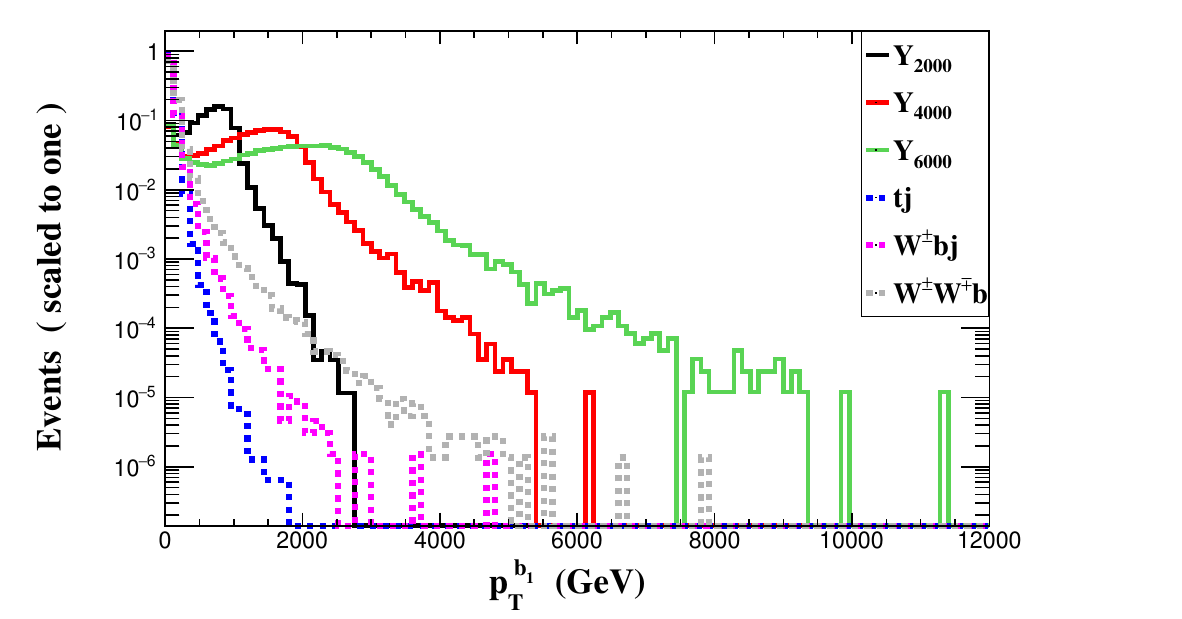}
		%\label{chutian1}%文中引用该图片代号
	\end{minipage}
	%\qquad
	\begin{minipage}{0.47\linewidth}
		\centering
		\includegraphics[width=1.1\linewidth]{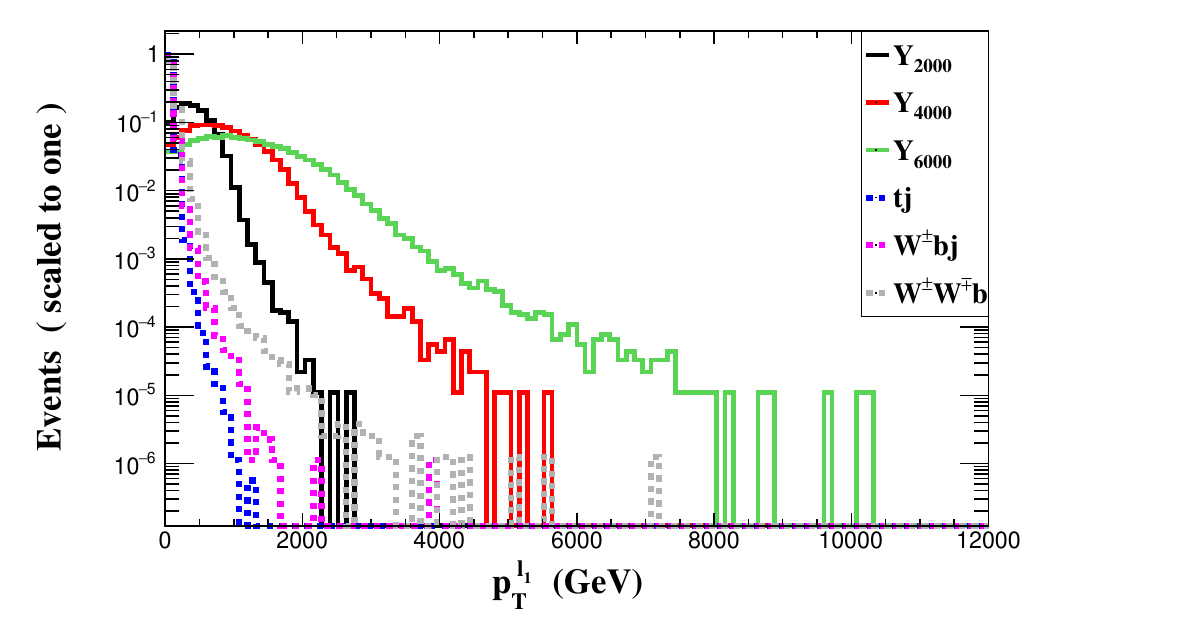}
	\end{minipage}
	%\qquad
	\begin{minipage}{0.47\linewidth}
		\centering
		\includegraphics[width=1.1\linewidth]{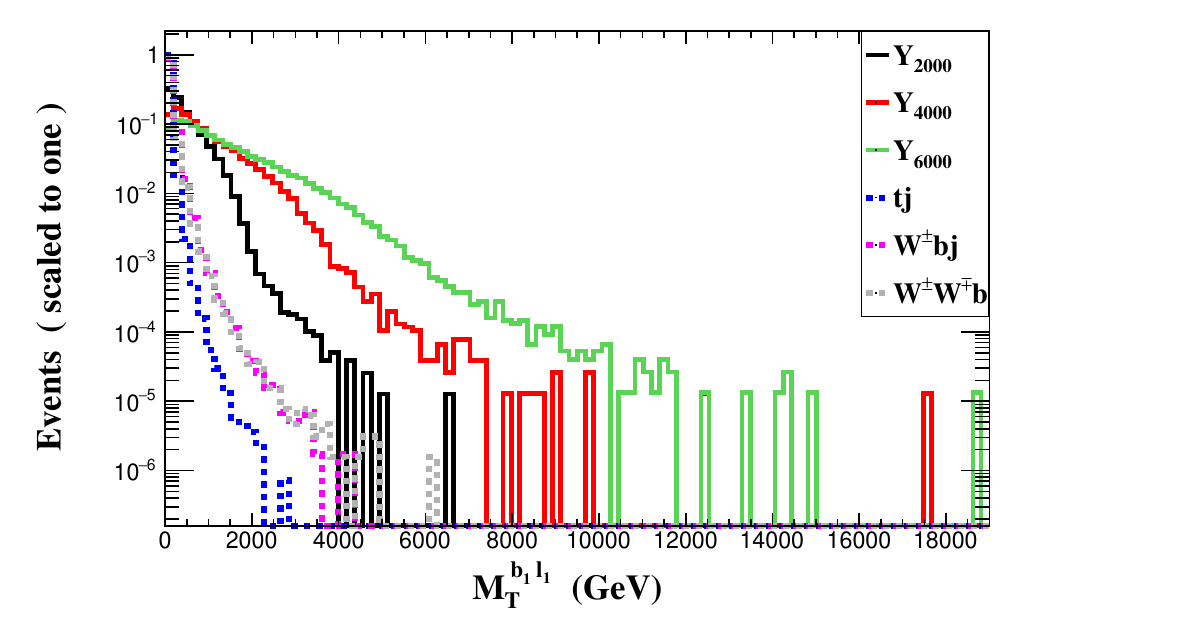}
	\end{minipage} 
    \begin{minipage}{0.47\linewidth}
		\includegraphics[width=1.1\linewidth]{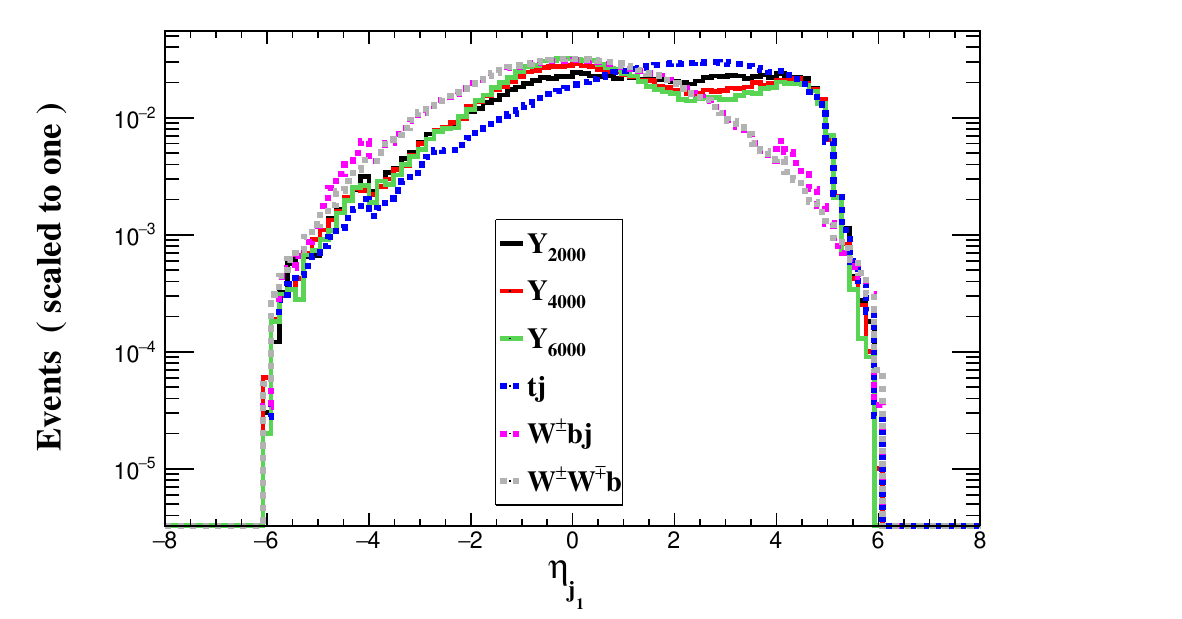}
    \end{minipage}    
 \begin{minipage}{0.47\linewidth}
		\includegraphics[width=1.1\linewidth]{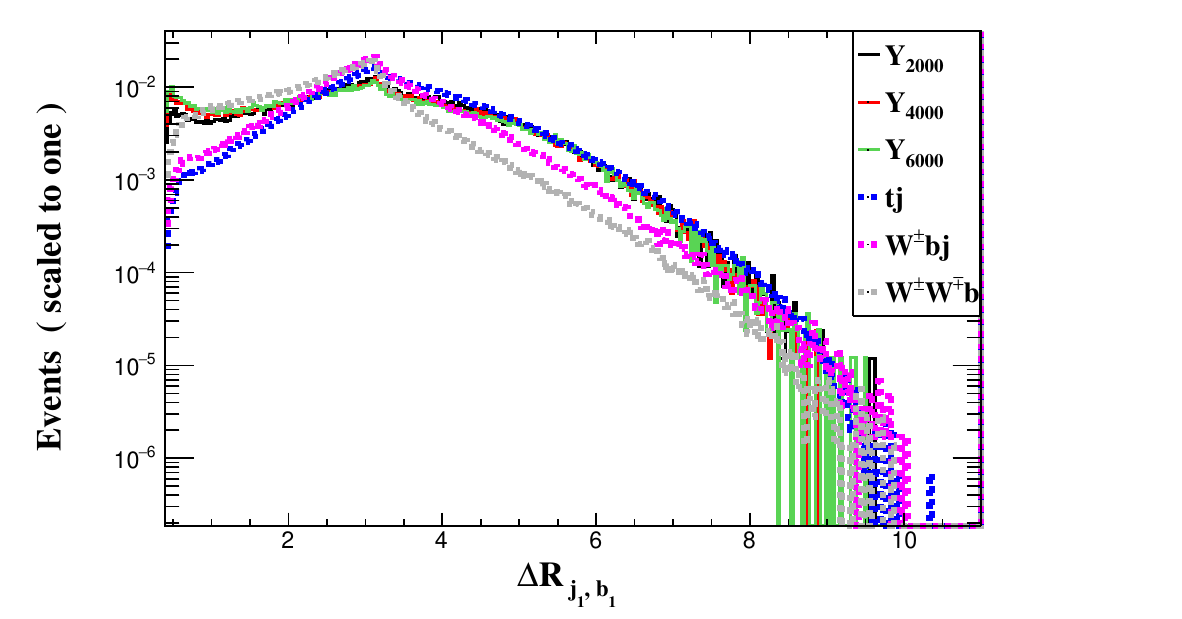}
		%\label{chutian1}%文中引用该图片代号
	\end{minipage}
	%\qquad
 \begin{minipage}{0.47\linewidth}
		\centering
		\includegraphics[width=1.1\linewidth]{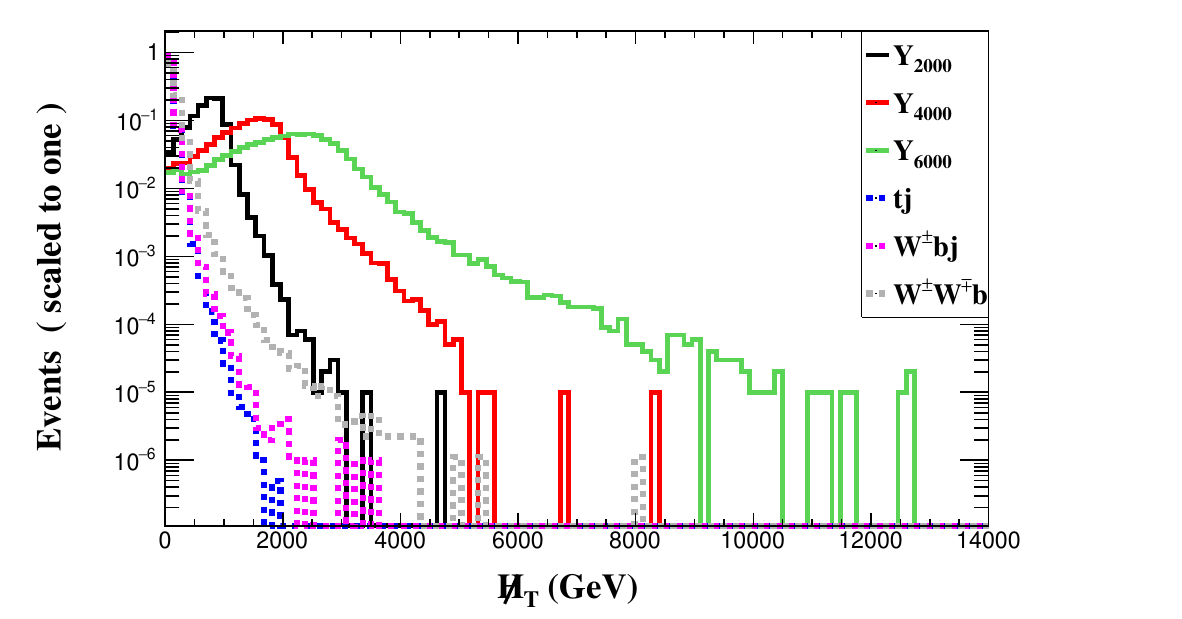}
	\end{minipage} 
    \begin{minipage}{0.47\linewidth}
		\includegraphics[width=1.1\linewidth]{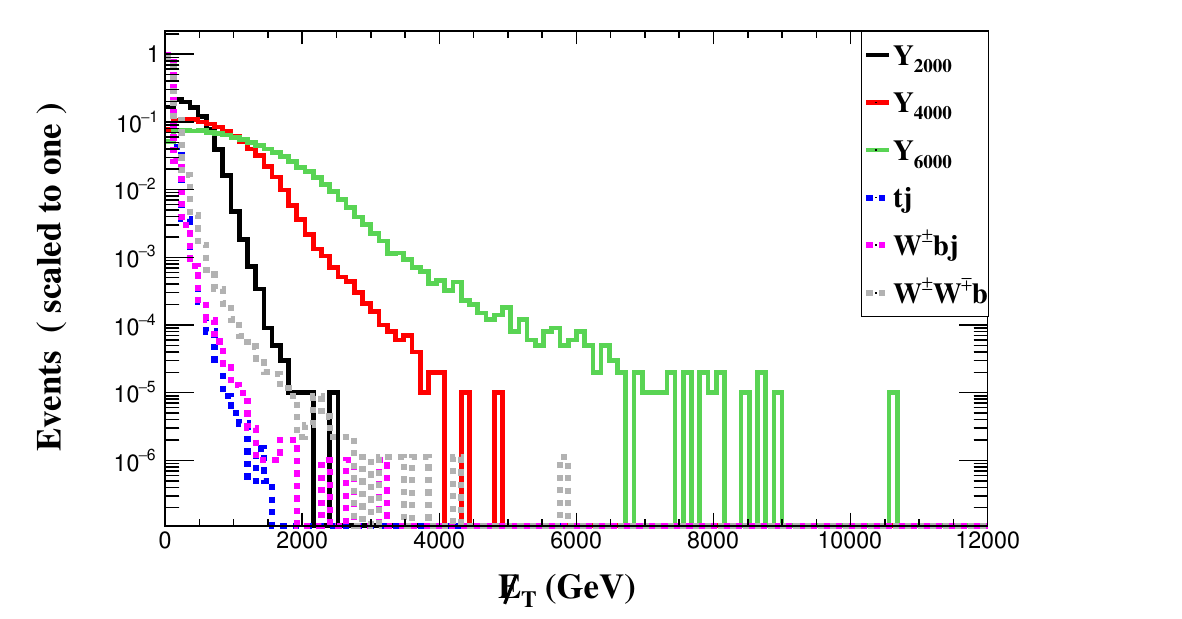}
    \end{minipage}    
	
	\caption{ Normalized distributions for the signals with $m_Y=$ 2000 GeV, 4000 GeV and 6000 GeV, and backgrounds at the FCC-hh.}% The conjugated processes have been included.} 
		\label{100fenbutu}
\end{figure*}

%信号与背景的区别

Here, we explore the anticipated signal of ${Y}$ in the context of the future 100 TeV FCC-hh. We commence the calculation from $m_{Y}=1500\text{ GeV}$ up to 15000 GeV and select benchmark points at intervals of 250 GeV with $\kappa_Y=0.1$. The figures in Figure~\ref{100fenbutu} portray normalized distributions for signal of $m_Y = 2000\text{ GeV}, 4000\text{ GeV}$ and 6000 GeV as well as and background processes, laying the groundwork for our distinctive selection criteria: 
\begin{itemize}
    \item {Cut-1}: $N_{l} = 1$, $N_{b} \ge 1$, $N_{j} \ge 1$;
    \item Cut-2: $p_{T}^{b_1} > 700 \text{ GeV}$ and $p_{T}^{l_1} > 220 \text{ GeV}$;
    %\item Cut-3: $M_{b_1 l_1} > 700 \text{ GeV}$;
    %\item Cut-4: $M_{b_1 l_1 j_1} > 1000 \text{ GeV}$;
    \item {Cut-3}: $M_{T}^{b_1 l_1} > 200 \text{ GeV}$;
    \item Cut-4: $2.4 < \eta _{j_1} < 4.8 $;
    \item Cut-5: $\Delta R_{j_1 b_1} > 3.0 $; 
    \item {Cut-6}: $\slashed{H}_{T} > 400 \text{ GeV}$ and $\slashed{E}_{T} > 150 \text{ GeV}$.
\end{itemize}
\begin{table}[ht]
	\centering
	\begin{threeparttable}
		\begin{tabular}{c|ccc|c|c|c}
			\hline
			Cuts       & $Y_{2000}$ (fb) & $Y_{4000}$ (fb) & $Y_{6000}$ (fb)&$tj$\,\, (fb)  &$W^\pm bj$\,\,(fb)    &$W^\pm W^\mp b$\,\,(fb)   \\ \hline
			Basic Cuts    & 155.58  & 30.65 & 9.06 & 182592    & 411730    & 573258  \\
			Cut 1         & 90.71   & 17.48 & 5.09 & 90355     & 163348    & 299942  \\
			Cut 2         & 33.82   & 12.18 & 3.77 & 1.19      & 76.17     & 580\\
			%Cut 3         & 33.77   & 12.15 & 3.77 & 1.19      & 73.70     & 475.90   \\
			%Cut 4         & 33.72   & 12.15 & 3.77 & 1.19      & 73.29     & 474     \\
			Cut 3         & 25.3    & 10.87 & 3.56 & 1         & 67.11     & 481     \\
			Cut 4         & 10.90   & 4.02  & 1.22 & 0.27      & 3.29      & 49.82  \\
                Cut 5         & 9.52    & 3.43  & 1.04 & 0.27      & 2.90      & 35.77  \\
                Cut 6         & 8.35    & 3.21  & 0.99 & 0.09      & 0.82      & 16  \\
			\hline
		\end{tabular}
	\end{threeparttable}
	\caption{\label{cutflow100} Cut flows of the signal with $\kappa_Y=0.1$ and backgrounds at the 100 TeV FCC-hh.}%, where the conjugate processes have been included.}
\end{table}
Compared to the 14 TeV case, an additional variable $\Delta R_{j_1 b_1}$ is introduced here. Upon analyzing the distributions of $\Delta R_{j_1 b_1}$, it is apparent that the signal tends to be more central than the backgrounds. Thus, we require $\Delta R_{j_1 b_1} > 3.0 $.
The signal efficiencies for $m_Y=2000\text{ GeV}$, $m_Y=4000\text{ GeV}$ and $m_Y=6000\text{ GeV}$ are approximately 5\%, 10\% and 11\%, respectively. Notably, there is a significant suppression in the background processes. Comprehensive cut flows are provided in Table \ref{cutflow100}. The exclusion capability and discovery potential are illustrated in the final row of Figure~\ref{14-liangdutu}. It is evident that the systematic uncertainty has a considerable impact on the results. Even with a 10\% systematic uncertainty, the parameter space will significantly shrink. Accounting for the 10\% systematic uncertainty, the $Y$ quark can be excluded within the correlated parameter space of $\kappa_{Y} \in [0.08, 0.5]$ and $m_{Y} \in [1500\text{ GeV}, 10080\text{ GeV}]$ at the highest design value of luminosity, $L=30\text{ ab}^{-1}$. 
%Additionally, the $Y$ state can be discovered within $\kappa_{Y} \in [0.088 , 0.5]$ and $m_{Y} \in [1000\text{ GeV}, 4624\text{ GeV}]$ at $L=30\text{ ab}^{-1}$.

\begin{figure*}[p!]
\vspace{-1cm}
\centering
	\begin{minipage}{0.48\linewidth}
		\centering
		\includegraphics[width=\linewidth]{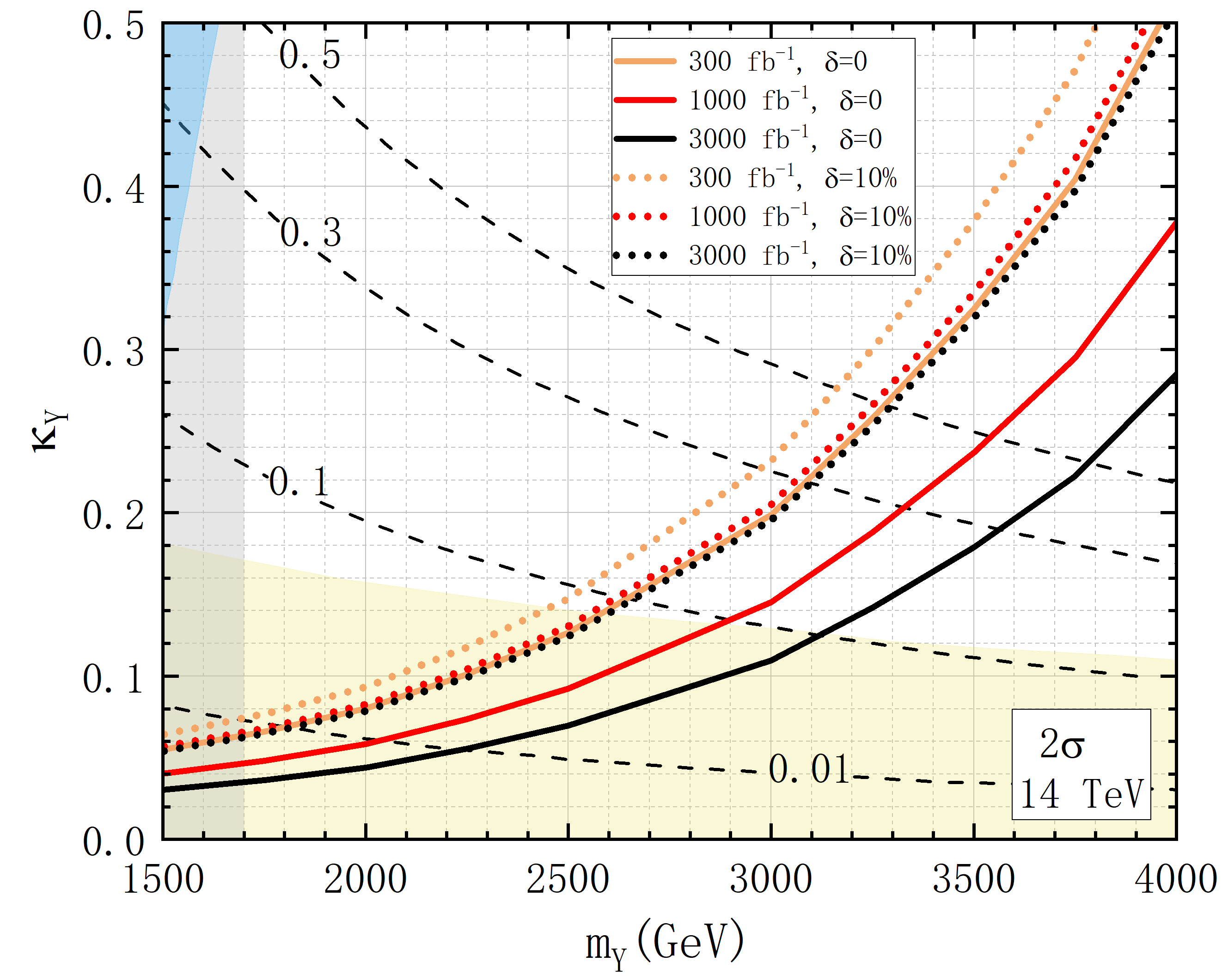}
	\end{minipage}
	%\qquad
	\begin{minipage}{0.48\linewidth}
		\centering
		\includegraphics[width=\linewidth]{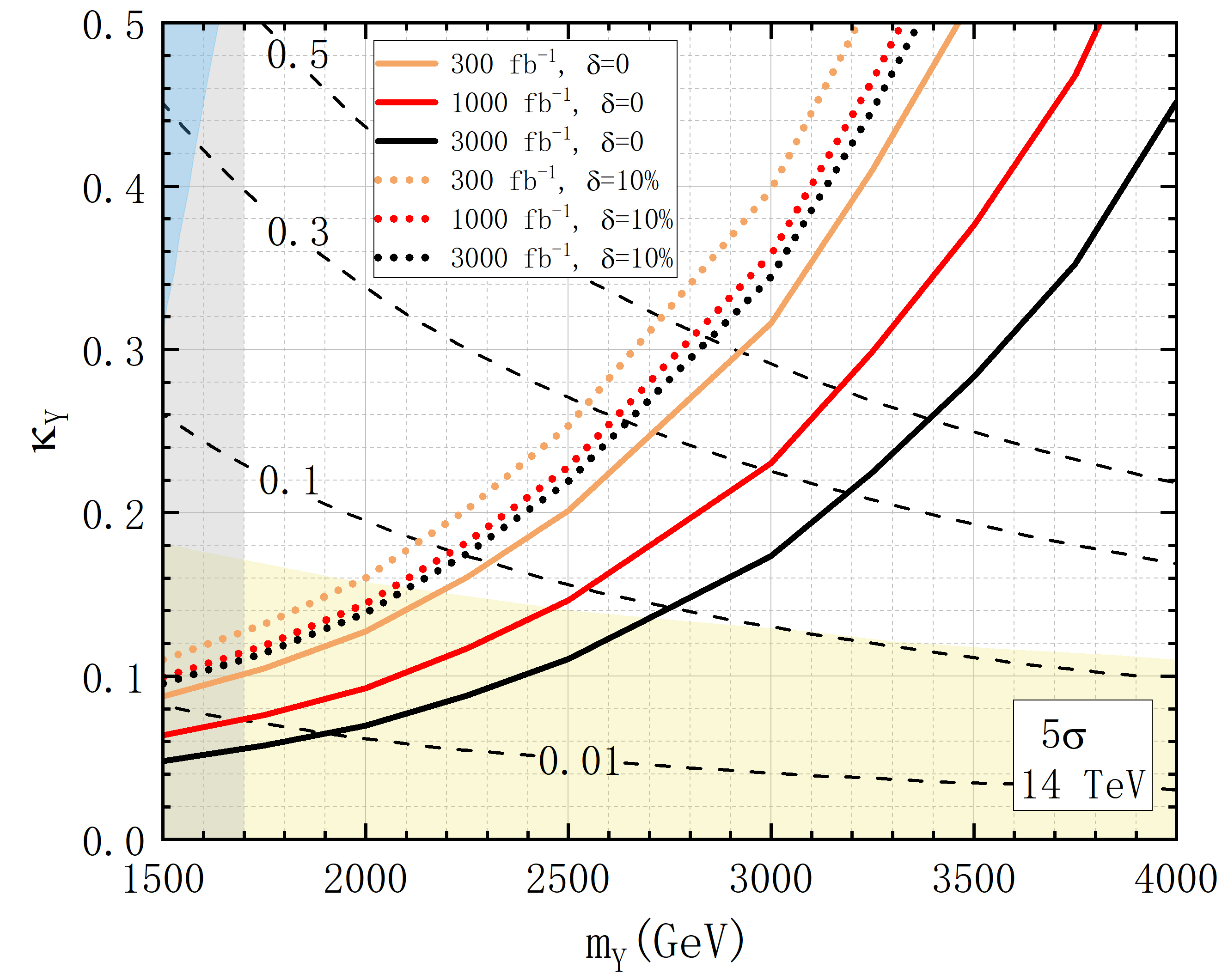}
	\end{minipage}

\centering
	\begin{minipage}{0.48\linewidth}
		\centering
		\includegraphics[width=\linewidth]{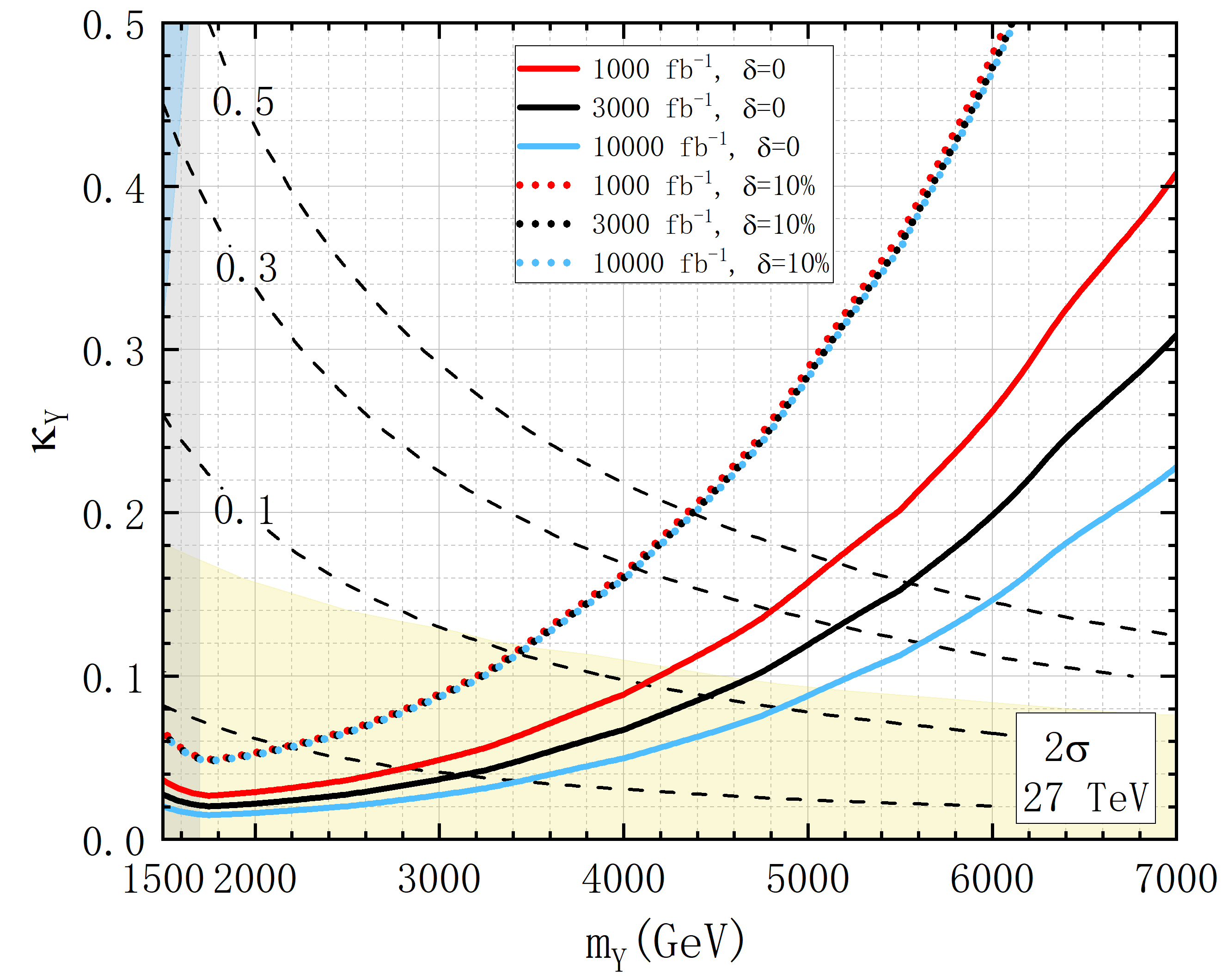}
	\end{minipage}
	%\qquad
	\begin{minipage}{0.48\linewidth}
		\centering
		\includegraphics[width=\linewidth]{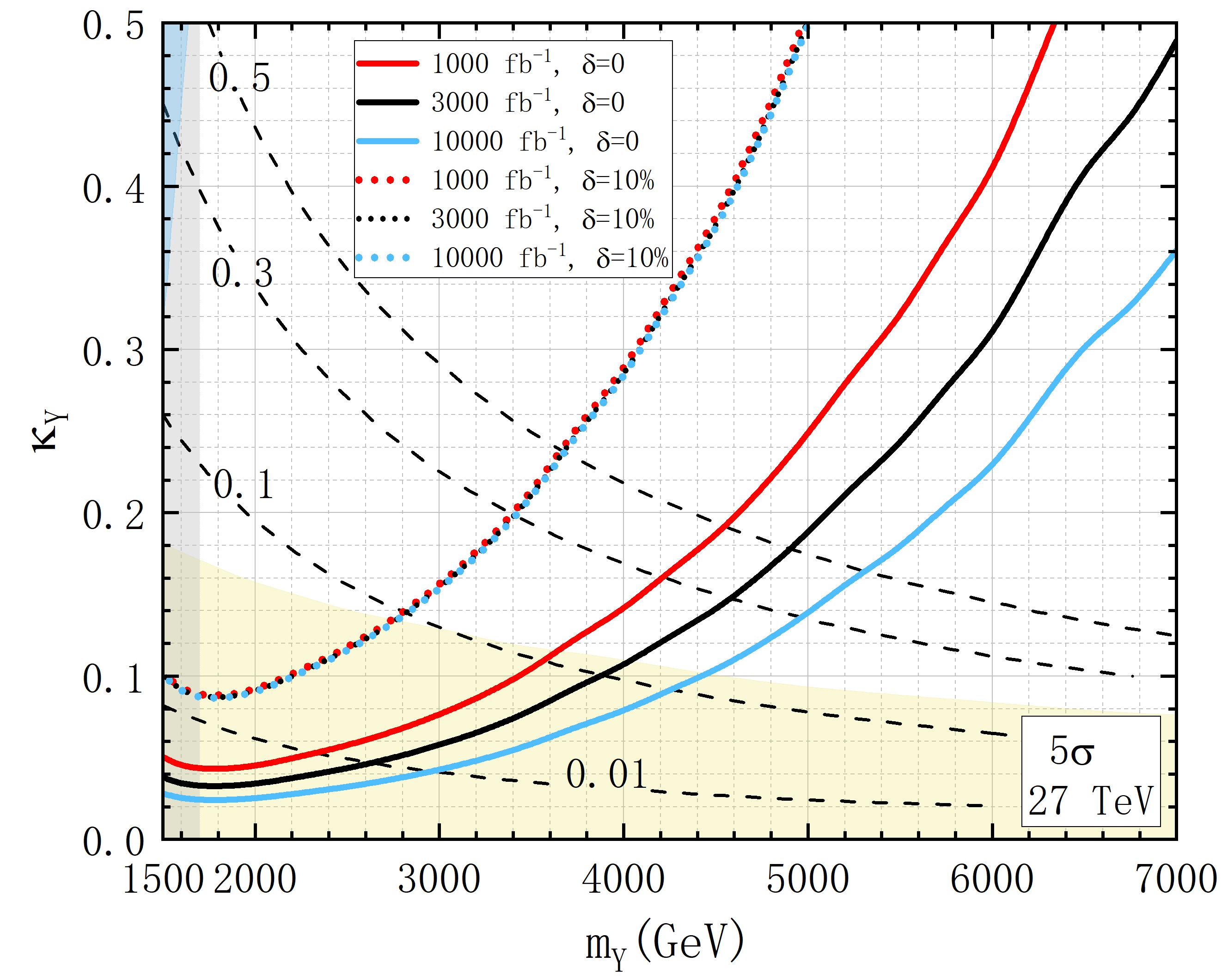}
	\end{minipage}

\centering
	\begin{minipage}{0.48\linewidth}
		\centering
		\includegraphics[width=\linewidth]{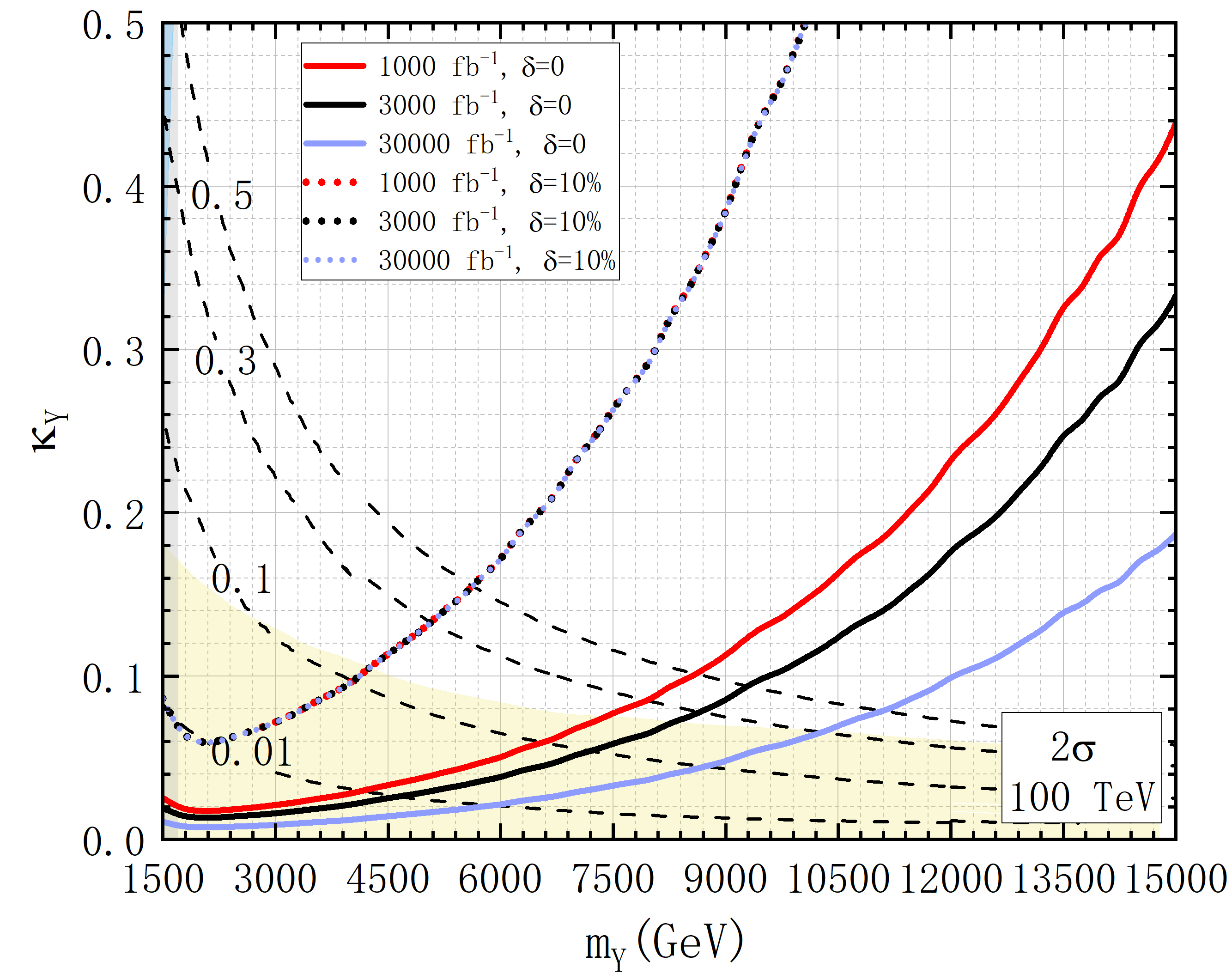}
	\end{minipage}
	%\qquad
	\begin{minipage}{0.48\linewidth}
		\centering		
            \includegraphics[width=\linewidth]{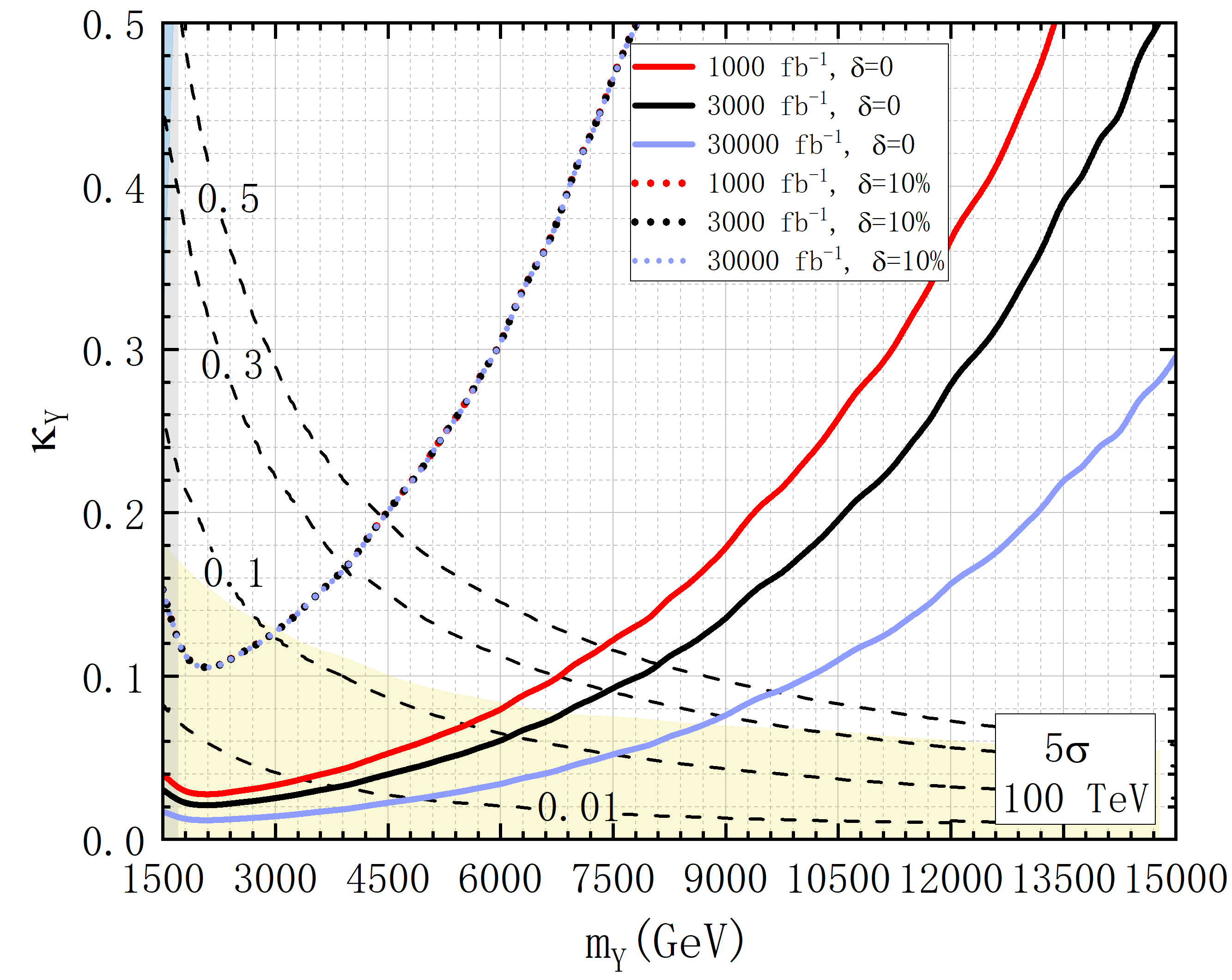}
	\end{minipage}
\caption{The exclusion capability ($\mathcal{Z}_{\text{excl}}=2$) and discovery potential ($\mathcal{Z}_{\text{disc}}=5$) for the $Y$ state at the LHC Run-III and HL-LHC,  $\sqrt{s}=27\text{ TeV}$ HE-LHC and $\sqrt{s}=100\text{ TeV}$ FCC-hh. Solid lines represent the ideal scenario without systematic uncertainty, the dotted lines represent the scenario with a 10\% systematic uncertainty. Dashed lines denote the contours of $\Gamma_Y/m_Y$. The blue (grey) shaded area indicates the exclusion region of the current LHC at $\sqrt{s}=$ 13 TeV with $L=$36.1 fb$^{-1}$ (140 fb$^{-1}$), as reported in Ref.~\cite{ATLAS:2018dyh} (Ref.~\cite{ATLAS:2023shy}). Meanwhile, the yellow shaded area denotes the allowed region for the oblique parameters $S, T$ and $U$, considering the current measurements in Ref.~\cite{ParticleDataGroup:2022pth}.
%Listing the exclusion capability and discovery potential for different $pp$ colliders in the same figure allows for convenient comparison between the results obtained from each center-of-mass energy.
}
\label{14-liangdutu}
\end{figure*}

\section{Summary}
\label{sec:summary}

\iffalse
\begin{table}[]
\begin{tabular}{ccccccc}
\hline
Colliders                        & $L/\text{fb}^{-1}$ & Uncertainty & \multicolumn{2}{c}{Exclusion}                   & \multicolumn{2}{c}{Discovery}                  \\ \hline
                                &             &             & $\kappa_{Y}$ & $m_Y$(GeV)         & $\kappa_{Y}$ & $m_Y$(GeV)        \\ \hline
   {LHC Run-III} & 300         & 10\%        & {[}0.21,0.47{]}              & {[}1000,4000{]}  & {[}0.072,0.5{]}            & {[}1000,2621{]} \\ \hline
{14 TeV HL-LHC} & 3000        & 10\%        & {[}0.17,0.37{]}              & {[}1000,3653{]}  & {[}0.047,0.5{]}              & {[}1000,3047{]} \\ \hline
{27 TeV HE-LHC} & 10000        & 10\%        & {[}0.029,0.5{]}              & {[}1000,4987{]}  & {[}0.051,0.5{]}              & {[}1000,3943{]} \\ \hline
{100 TeV FCC-hh} & 30000        & 10\%        & {[}0.051,0.5{]}              & {[}1000,6610{]}  & {[}0.088,0.5{]}               & {[}1000,4624{]} \\ \hline
\end{tabular}
    \caption{Summary for 2$\sigma$ exclusion limits and 5$\sigma$ signal discoveries at the LHC Run-III and HL-LHC,  $\sqrt{s}=27\text{ TeV}$ HE-LHC and $\sqrt{s}=100\text{ TeV}$ FCC-hh. }
    \label{zongjie}
\end{table}   
\fi

In a simplified model, we have investigated the single production of a doublet VLQ denoted by $Y$ in the $W b$ decay channel at the the $\sqrt s=14$ TeV HL-LHC, $\sqrt s=27$ TeV HE-LHC and $\sqrt s=100$ TeV FCC-hh, following its production via  $pp\to Yj$, with the $W$ decaying into electrons and muons plus their respective neutrinos. 
%在一个简化的模型中，我们研究了具有类向量粒子Y的W b道路上Y的单个产生，其中我们使用了5FS产生事件。我们对信号和背景进行了详细的探测器模拟，并显示了在不同对撞机上对VLQ-Y的排除和发现能力。在HL-LHC、HE-LHC和FCC-hh上的VLQ-Y的排除和发现能力。
We have performed a detector level simulation for the signal and relevant SM backgrounds. Considering a systematic uncertainty of 10\% with an integrated luminosity of 3000 fb$^{-1}$,
%\sout{the exclusion and discovery capabilities, as displayed in Table~VI, can be described as follows:}
{we describe the exclusion and discovery capabilities as follows:}
(1) The HL-LHC can exclude (discover) the region of $\kappa_{Y} \in [0.05,0.5]$ $([0.09, 0.5])$ and $m_{Y} \in [1500\text{ GeV}, 3970\text{ GeV}]$ $([1500\text{ GeV}, 3360\text{ GeV}])$; 
(2) The HE-LHC can exclude (discover) the region of $\kappa_Y \in [0.06 , 0.5]$ $([0.1 , 0.5])$ and $m_{Y} \in [1500\text{ GeV}, 6090\text{ GeV}]$ $([1500\text{ GeV}, 5000\text{ GeV}])$; 
(3) The FCC-hh can exclude (discover) the region of $\kappa_Y \in [0.08 , 0.5]$ $([0.14 , 0.5])$ and $m_{Y} \in [1500\text{ GeV}, 
10080\text{ GeV}]$ $([1500\text{ GeV}, 7800\text{ GeV}])$.
All information of $Y$ exclusion and discovery can be clearly seen in Figure~\ref{14-liangdutu}. 

If the mass of $Y$ is notably greater than that of its decay products and much heavier than the particles originating from background processes, the signal exhibits more distinct characteristics compared to backgrounds. However, as the mass of $Y$ increases, the cross section of the signal decreases. Therefore, there is a balance between these two aspects, and we observe an optimal discovery reach depicted as a curve bulge in Figure~\ref{14-liangdutu}.
%The upper limits of $m_Y$ and $\kappa_Y$ are dependent on the parameter range we choose. 
%For example, if a larger $\kappa_Y$ greater than 0.5 is considered, provided it does not exceed the perturbative limit, the upper limit for $\kappa_Y$ will be relaxed.

Furthermore, we highlight that the stringent constraint on the VLQ $Y$, derived from the $Y$ pair production search with BR($Y\to W^- b$) $=1$, imposes $m_Y>1700\text{ GeV}$. In this context, we reassess the potential of LHC Run-III to explore the VLQ $Y$, revealing that the associated parameter regions of $\kappa_{Y} \in [0.06, 0.5]$ $([0.11, 0.5])$ and $m_{Y} \in [1500\text{ GeV}, 3800\text{ GeV}]$ $([1500\text{ GeV}, 3200\text{ GeV}])$ can be excluded (discovered) based on LHC Run-III luminosity. 
We foresee that our investigation will spur complementary explorations for a potential $Y$ quark at forthcoming $pp$ colliders.
%我们对信号和相关的标准模型背景进行了完整的模拟。总结起来，以最高的积分亮度来描述排除能力如下：(1) HL-LHC可以在积分亮度为3ab^(-1)时排除κY ∈ [0.05, 0.5]和mY ∈ [1500, 3725]的相关区域； (2) HE-LHC可以在积分亮度为3ab^(-1)时排除κ ∈ [0.03, 0.5]和mY ∈ [1500, 5891]的相关区域；(3) FCC-hh可以排除κ ∈ [0.03, 0.5]，κY ∈ [0.02, 0.5]和mY ∈ [1500, 10992]的相关区域。可以看出，随着质心能量的增加，排除能力明显增强，发现能力也具有类似的特点。我们期望本研究可以推动未来pp对撞机对可能的双重态Y夸克进行补充性搜索。

\section*{Acknowledgements}
This work of LS, YY and BY is supported by the Natural Science Foundation of Henan Province under Grant No. 232300421217, the National Research Project Cultivation Foundation of Henan Normal University under Grant No. 2021PL10, the China Scholarship Council under Grant No. 202208410277 and also powered by the High Performance Computing Center of Henan Normal University. 
The work of SM is supported in part through the NExT Institute, the 
Knut and Alice Wallenberg
Foundation under the Grant No. KAW 2017.0100 (SHIFT) and the STFC Consolidated Grant No. ST/
L000296/1.
\clearpage

\clearpage

%\input{sections/section01.tex}  
%\input{sections/section02.tex}
%\input{sections/section03.tex}  
%\input{sections/section04.tex}
%\input{sections/acknowledgements.tex}
%\clearpage
%\input{sections/appendix1.tex} 
%\clearpage
\bibliographystyle{utphys}
\bibliography{cit.bib}
\end{document}